\documentclass[12pt]{article}

\usepackage{color}
\definecolor{darkblue}{rgb}{0.1,0.1,.7}
\usepackage[colorlinks, linkcolor=darkblue, citecolor=darkblue, urlcolor=darkblue, linktocpage]{hyperref}
\usepackage[]{amsmath}
\usepackage[]{graphicx}
\usepackage[]{latexsym}
\usepackage{slashed,graphicx,color,amsmath,amssymb}
\usepackage[mathscr]{eucal}
\usepackage{mathrsfs}
\usepackage{geometry}
\usepackage[margin=10pt,font=small,labelfont=bf]{caption}
\usepackage{amscd}
\usepackage{bm}
\usepackage{xcolor}
\usepackage[thicklines]{cancel}
\usepackage{upgreek}
\usepackage[square, comma, sort&compress,numbers]{natbib}
\usepackage[all,cmtip]{xy}
\usepackage[color=cyan!30!white,linecolor=red,textsize=footnotesize]{todonotes}
\numberwithin{equation}{section}
 \makeatletter
 \g@addto@macro\bfseries{\boldmath}
 \makeatother

\begin{document}
\vspace*{-.6in} \thispagestyle{empty}
\vspace{.2in} {\large
\begin{center}
{\bf Integrable Crosscap States:\\
From Spin Chains to 1D Bose Gas}
\end{center}}
\vspace{.2in}
\vspace{.1in}
\renewcommand{\thefootnote}{\fnsymbol{footnote}}
\begin{center}
{Miao He$^{a,b}$\footnotemark[1]\quad Yunfeng Jiang$^{a,b}$\footnotemark[2]
\vspace{.2in}\\
\textit{$^{a}$School of Physics, Southeast University, Nanjing 211189, China}\\
\textit{$^{b}$Shing-Tung Yau Center, Southeast University, Nanjing 210096, China}}\\
\end{center}
\footnotetext[1]{E-mail: hemiao@seu.edu.cn}
\footnotetext[2]{E-mail: jinagyf2008@seu.edu.cn}
\vspace{.3in}

\begin{abstract}
The notion of a crosscap state, a special conformal boundary state first defined in 2d CFT, was recently generalized to 2d massive integrable quantum field theories and integrable spin chains. It has been shown that the crosscap states preserve integrability. In this work, we first generalize this notion to the Lieb-Liniger model, which is a prototype of integrable non-relativistic many-body systems. We then show that the defined crosscap state preserves integrability. We derive the exact overlap formula of the crosscap state and the on-shell Bethe states. As a byproduct, we prove the conjectured overlap formula for integrable spin chains rigorously by coordinate Bethe ansatz. It turns out that the overlap formula for both models take the same form as a ratio of Gaudin-like determinants with a trivial prefactor. Finally we study quench dynamics of the crosscap state, which turns out to be surprisingly simple. The stationary density distribution is simply a constant. We also derive the analytic formula for dynamical correlation functions in the Tonks-Girardeau limit.
\end{abstract}

\vskip 1cm \hspace{0.7cm}

\newpage

\setcounter{page}{1}
\begingroup
\hypersetup{linkcolor=black}
\tableofcontents
\endgroup
\renewcommand{\thefootnote}{\arabic{footnote}}
\section{Introduction}
\label{sec:1}
An integrable quantum field theory (IQFT) in 1+1 dimension has infinitely many local conserved charges. In the presence of boundaries, some of these charges are no longer conserved. Nevertheless, there are special boundary conditions which preserves an infinite subset of the conserved charges. These boundary conditions are called integrable. For a Lorentz invariant theory, one can equivalently place the boundary in the temporal direction, in which case the boundary condition becomes a special state in the Hilbert space called an integrable boundary state~\cite{Ghoshal:1993tm}. A characteristic feature of integrable boundary states is that they are annihilated by infinitely many odd charges of the model. In recent years, it has become clear that the notion of integrable boundary states can be generalized to a broader class of theories such as integrable lattice models~\cite{Piroli:2017sei,Pozsgay:2018dzs}.
\par
Interests on integrable boundary states stem from both statistical mechanics and AdS/CFT correspondence. In statistical mechanics, these states can serve as initial states for the investigation of out-of-equilibrium dynamics~\cite{Caux:2013ra,Sotiriadis:2013fca,PhysRevLett.113.117202,PhysRevLett.113.117203,Essler:2016ufo}. Due to their integrability, one can have more analytic control for the calculations. In AdS/CFT correspondence, it turns out that various kinds of correlation functions in $\mathcal{N}=4$ SYM theory and ABJM theory at weak coupling can be computed by the overlap of an on-shell Bethe state and an integrable boundary state. These include the one-point functions in defect CFT~\cite{deLeeuw:2015hxa,Buhl-Mortensen:2015gfd,deLeeuw:2016umh,DeLeeuw:2018cal,Kristjansen:2020mhn,Kristjansen:2021abc,Gombor:2022aqj}, three-point functions of two giant gravitons and one non-BPS single-trace operator~\cite{Jiang:2019xdz,Jiang:2019zig,Yang:2021hrl}, and correlation functions of involving circular Wilson loops~\cite{TA1,TA2} and 't Hooft loops \cite{Kristjansen:2023ysz}. In all these cases, the exact overlap formulae play an important role. For integrable boundary states, only the Bethe states with parity even rapidities lead to non-vanishing overlaps. Moreover, the overlap formula has very nice analytic structure~\cite{Pozsgay_2014,Brockmann_2014,Brockmann_2014_0,Brockmann_2014_1,Foda:2015nfk,Pozsgay_2018}. In all known cases, it can be written as a the product of a prefactor and a ratio of Gaudin-like determinants. The former is state dependent while the latter is universal and only depends on the symmetry. The exact formulae have been proven in a number of cases using both the coordinate Bethe ansatz~\cite{Jiang:2020sdw,Chen:2020xel} and algebraic Bethe ansatz~\cite{Gombor:2021uxz,Gombor:2021hmj,Pozsgay_2014,Brockmann_2014,Brockmann_2014_0,Brockmann_2014_1,Foda:2015nfk,Pozsgay_2018}, while for other cases they remain conjectures with extensive numerical evidence.
\par
Very recently, a new type of integrable boundary states called crosscap states have been investigated. These states first arise in 2d CFT, which are special conformal boundary states~\cite{Ishibashi:1988kg}. Geometrically, they correspond to non-orientable surfaces such as $\mathbb{RP}^2$ and the Klein bottle, which cannot be described by a local boundary condition. In~\cite{Caetano:2021dbh}, by generalizing the geometric intuition from CFT, the authors defined crosscap states for 2d massive IQFTs and the integrable spin chains. Remarkably, they discovered that crosscap states as they defined are integrable. The crosscap states have several new features which make them rather special. First of all, most known integrable boundary states such as the two-site states and matrix product states are short-range entangled while crosscap states are long-range entangled by construction. In addition, the overlap formula for the crosscap state and an on-shell Bethe state has a trivial prefactor~\cite{Caetano:2021dbh}, and is given simply by the ratio of Gaudin-like determinants. In this sense, crosscap states are probably the `cleanest' integrable boundary states. The unique features of the crosscap states are intimately related to their geometric origin. Therefore we expect it should be possible to define such states for a wide class of models. Indeed, generalizations to $\mathfrak{gl}(N)$ spin chains~\cite{Gombor:2022deb} and classical sigma models~\cite{Gombor:2022gdj} have been studied recently. 
\par
Apart from relativistic IQFTs and spin chains, there is yet another important class of integrable models which are continuous but non-relativistic. The prototype of these models is the Lieb-Liniger model~\cite{PhysRev.130.1605}, which describes one dimensional bosonic particles interacting with a pairwise $\delta$-function potential. Apart from theoretical interests, the Lieb-Liniger model has direct relevance to cold atom experiments (see for examples the reviews \cite{GuanReview2,Batchelor:2015osa,GuanReview1}). It is particularly interesting to compute its dynamical observables since they can be measured in the laboratory. 
However, so far the only known integrable boundary state for the Lieb-Liniger model is the so-called Bose-Einstein Condensate (BEC) state. It was found that the overlap of the BEC state and on-shell Bethe states obey parity even selection rules and the overlap formula exhibit the same structure as spin chains~\cite{PhysRevA.89.033601,PhysRevA.89.013609,Chen:2020xel}. 
Since both for IQFTs and spin chains, one can construct many integrable boundary states, it is a natural question whether we can construct more integrable boundary states for the Lieb-Liniger model. In view of its close relation to experiments, such integrable boundary states might be even more interesting. The crosscap state seems to be a natural candidate, as its geometric origin gives us a clear guidance for its construction. As we will show below this indeed turns out to be the case.\par

Now let us sketch our strategy for the construction. The Lieb-Liniger model sits somewhere between IQFTs and spin chains. On the one hand, it can be obtained as the non-relativistic limit of the sinh-Gordon model~\cite{Kormos:2009eqa,Kormos:2009yp}; on the other hand, it corresponds to special continuum limits of certain spin chain models~\cite{Golzer_1987,Pozsgay:2011ec}. Therefore we can start with crosscap states in either IQFTs or spin chains and then take the proper limit. It turns out that the second option is more feasible. We will comment on its relation to the first option later. We consider two methods to obtain crosscap state in the Lieb-Liniger model from spin chains: the first one involves taking a special continuum limit of the XXZ spin chain~\cite{Golzer_1987}, while the second one involves discretizing the Lieb-Liniger model as a generalized XXX spin chain~\cite{Izergin:1982ry}. It turns out that the two methods lead to the same result, given by
\begin{align}
\label{eq:CCstart}
|\mathcal{C}\rangle=\exp\left(\int_{0}^{\ell/2}\mathrm{d}x\,\Phi^{\dagger}(x)\Phi^{\dagger}(x+\ell/2)\right)|\Omega\rangle\,,
\end{align} 
where $\Phi^{\dagger}(x)$ is the bosonic operator which creates a particle at position $x$. The geometric meaning of \eqref{eq:CCstart} is quite clear --- particles are created in pairs at antipodal points. As a result, this state is long-range entangled by construction. One might notice that this state is similar to the integrable boundary states constructed by Ghoshal and Zamolodhikov in IQFT \cite{Ghoshal:1993tm} where one replaces $\Phi^{\dagger}(x)\Phi^{\dagger}(x+\tfrac{\ell}{2})$ by $K^{ab}(\theta)A_a^{\dagger}(\theta)A_b^{\dagger}(-\theta)$. However, we want to emphasis two important differences. First, the Faddeev-Zamolodchikov operator $A_a^{\dagger}(\theta)$ is a creation operator in \emph{momentum space} while $\Phi^{\dagger}(x)$ is the creation operator in \emph{position space}. Second, the Ghoshal-Zamolodchikov construction describes boundary states in the infinite volume while for the crosscap state finite volume is necessary to define antipodal points.\par 

The rest of this paper is organized as follows. In section~\ref{sec:2}, we generalize the proof of integrability of crosscap states to anisotropic inhomogeneous Heisenberg spin chains, and derive the exact overlap formula using the coordinate Bethe ansatz. This section can be seen as a useful technical preparation for similar calculations in the Lieb-Liniger model. In section~\ref{sec:3}, we propose the crosscap state in the Lieb-Liniger model by taking the continuum limit from spin chain crosscap state. We prove its integrability and derive the exact overlap formula of the crosscap state and an on-shell Bethe state. The dynamical correlation function in crosscap states are studied in section~\ref{sec:4}. We conclude and discuss future directions in section~\ref{sec:5}. Some details of the calculations are given in the appendices.

\section{Crosscap states of integrable spin chains}
\label{sec:2}
The crosscap state is constructed by identifying the states at antipodal sites. For XXX spin chain, the entangled pair of states at sites $j$ and $j+\frac{L}{2}$ is 
\begin{align}
|c\rangle\!\rangle_{j}=&\left(1+S_{j}^{+}S_{j+L/2}^{+}\right)|\!\downarrow\rangle_{j} \otimes|\!\downarrow\rangle_{j+L/2}.
\end{align}
where we denote the generators of the SU(2) algebra at site-$j$ by $S_j^{\pm}$ and $S_j^z$. The crosscap state is then defined by taking the tensor product of such entangled pairs
\begin{align}
\label{Heisenberg-crosscap}
|\mathcal{C}\rangle_{\text{SU}(2)} \equiv&\prod_{j=1}^{L/2}\left(|c\rangle\!\rangle_{j}\right)^{\otimes}=\prod_{j=1}^{L/2}\left(1+S_{j}^{+}S_{j+\frac{L}{2}}^{+}\right)|\Omega\rangle\,,
\end{align}
where the $|\Omega\rangle=|\downarrow^L\rangle$ is the pseudovacuum.
\par
The crosscap state can also be defined for the non-compact $SL(2,R)$ spin chain~\cite{Caetano:2021dbh}. The main difference is that more than one particles can be excited on the same site. In this case, the entangled pair at sites $j$ and $j+\frac{L}{2}$ reads
\begin{align}
|c\rangle\!\rangle_{j}=&\sum_{n=0}^{\infty}\frac{1}{n!}\left(S_{j}^{+}S_{j+L/2}^{+}\right)^{n}|0\rangle_{j} \otimes|0\rangle_{j+L/2},
\end{align}
where $|0\rangle$ represents the lowest-weight state of $SL(2,R)$. The crosscap state is given by
\begin{align}
\label{SL-crosscap}
|\mathcal{C}\rangle_{SL(2)} \equiv \prod_{j=1}^{L/2}\left(|c\rangle\!\rangle_{j}\right)^{\otimes}=\prod_{j=1}^{L/2}\left[\sum_{n=0}^{\infty}\frac{1}{n!}\left(S_{j}^{+}S_{j+L/2}^{+}\right)^{n}\right]|\Omega\rangle.
\end{align} 
Both \eqref{Heisenberg-crosscap} and \eqref{SL-crosscap} has been studied in~\cite{Caetano:2021dbh}. 
\par
In this section, we derive two new results which are useful for the Lieb-Liniger model. One is defining crosscap states for  the inhomogeneous XXZ spin chain and proving their integrability. This constitutes a slight generalization of the results in~\cite{Caetano:2021dbh}. The other is deriving the exact overlaps between crosscap states and on-shell Bethe states for both compact and non-compact spin chains. The overlap formula was first conjectured in~\cite{Caetano:2021dbh} and later proven in \cite{Gombor:2022deb} using algebraic Bethe ansatz for XXX spin chain (as a special case of $\mathfrak{gl}(N)$ spin chain). 
Here we give an alternative proof using CBA, which works for both compact and non-compact spin chains and can be generalized to the Lieb-Liniger model.

\subsection{Integrability of the crosscap state}
\label{sec:2.1}
The integrable boundary states $|\Psi_0\rangle$ are the states which are annihilated by the odd charges 
\begin{align}
Q_{2n+1}|\Psi_0\rangle=0.
\end{align}
For spin chains, it is more convenient to work with the transfer matrix, which is the generating functional of the conserved charges
\begin{align}
T(\lambda)= \exp \left(\sum_{n=1}^{\infty} \frac{\lambda^{n}}{n !} Q_{n+1}\right)\,.
\end{align} 
In~\cite{Piroli:2017sei,Pozsgay:2018dzs}, the authors propose to define integrable boundary states as the states satisfying
\begin{align}
\label{Integrable-condition}
T(\lambda)|\Psi_0\rangle=T(-\lambda)|\Psi_0\rangle.
\end{align}
We shall adopt this definition here.
\par
\paragraph{Inhomogeneous XXZ chain} Let us consider the inhomogeneous XXZ spin chain defined by the $R$-matrix\begin{align}
R(\lambda) & =\left(\begin{array}{llll}
\sinh(\lambda+\eta) & & & \\
& \sinh(\lambda) & \sinh(\eta) & \\
& \sinh(\eta) & \sinh(\lambda) & \\
& & & \sinh(\lambda+\eta)
\end{array}\right).
\end{align}
For later convenience, we can also write it in the tensor product form
\begin{align}
R(\lambda)=&\frac{1}{2}\Big(\sinh(\lambda+\eta)+\sinh(\lambda)\Big)\textbf{1}\otimes\textbf{1}+\Big(\sinh(\lambda+\eta)-\sinh(\lambda)\Big)S^{z}\otimes\sigma^{z}\nonumber\\
&+\sinh(\eta)\Big(S^{+}\otimes\sigma^{-}+S^{-}\otimes\sigma^{+}\Big).
\end{align}
where $\sigma^{\alpha}$ are Pauli matrices and $S^{\alpha}=\sigma^{\alpha}/2$. The Lax operator at site-$j$ is defined as
\begin{align}
L_{j}(\lambda)=R_{aj}\left(\lambda-\xi_{j}-\eta/2\right),
\end{align}
where $\xi_{j}$ is the inhomogeneity. The inhomogeneous XXZ spin chain is defined by the following transfer matrix
\begin{align}
T_{\text{XXZ}}(u)=\text{tr}_a\left(L_{1}(\lambda)\ldots L_{L}(\lambda)\right).
\end{align}
\paragraph{Crosscap state and integrability} For the inhomogeneous XXZ spin chain, we define the crosscap state to be the same state \eqref{Heisenberg-crosscap}. We now prove its integrability. Using the relation
\begin{align}
S^{\pm}_{j}|c\rangle\!\rangle_{j}=S^{\mp}_{j+L/2}|c\rangle\!\rangle_{j}\,,
\end{align}
we find that
\begin{align}
\label{lax-condition}
\sigma_{2}L_{j}(\lambda)\sigma_{2}|c\rangle\!\rangle_{j}=-L_{j+L/2}(-\lambda)|c\rangle\!\rangle_{j},
\end{align}
if the inhomogeneities at site $j$ and $j+L/2$ are also identified as~\cite{Ekman:2022kpx}
\begin{align}
\label{paired-homogeneities}
\xi_{j+L/2}=-\xi_{j},\quad j=1,2,...,L/2\,.
\end{align}
By employing~\eqref{lax-condition}, the action of the transfer matrix on the crosscap state reads
\begin{align*}
T_{\text{XXZ}}(\lambda)|\mathcal{C}\rangle & =\text{tr}_a\left[\left(L_{1}(\lambda) \cdots L_{L/2}(\lambda)\right)\left(L_{L/2+1}(\lambda) \cdots L_{L}(\lambda)\right)\right]|\mathcal{C}\rangle \nonumber\\
& =(-1)^{L/2}\text{tr}_a\left[\left(L_{1}(\lambda) \cdots L_{L/2}(\lambda)\right)\left(\sigma_{2} L_{1}(-\lambda) \cdots L_{L/2}(-\lambda) \sigma_{2}\right)\right] |\mathcal{C}\rangle \nonumber\\
& =(-1)^{L/2}\text{tr}_a\left[\left( L_{1}(-\lambda) \cdots L_{L/2}(-\lambda) \right)\left(\sigma_{2}L_{1}(\lambda) \cdots L_{L/2}(\lambda)\sigma_{2}\right)\right]|\mathcal{C}\rangle \nonumber\\
& =(-1)^{L}\text{tr}_a\left[\left( L_{1}(-\lambda) \cdots L_{L/2}(-\lambda) \right)\left(L_{L/2+1}(-\lambda) \cdots L_{L}(-\lambda)\right)\right]|\mathcal{C}\rangle \nonumber\\
&=T_{\text{XXZ}}(-\lambda)|\mathcal{C}\rangle. 
\end{align*}
Therefore the crosscap state is integrable if the inhomogeneities at antipodal sites have paired structure~\eqref{paired-homogeneities}. 


\subsection{Overlaps formula for compact chains}
\label{sec:2.2}
Owing to the condition \eqref{Integrable-condition}, the overlap of an on-shell Bethe state $|\boldsymbol{\lambda}_N\rangle$ and an integrable boundary state $|\Psi_0\rangle$ is only non-zero if the Bethe roots are parity even, namely $\{\boldsymbol{\lambda}_N\}=\{-\boldsymbol{\lambda}_N\}$. For $N$ being even, this implies
\begin{align}
\label{eq:paired}
\{\boldsymbol{\lambda}_N\}=\{\lambda_1,-\lambda_1,\lambda_2,-\lambda_2\ldots,\lambda_{N/2},-\lambda_{N/2}\}\,.
\end{align}
For parity even Bethe roots, it is known that the overlap take the following form\footnote{For some boundary states such as matrix product states with higher bond dimensions, the prefactor can take a more complicated form.}
\begin{align}
\label{overlaps}
\frac{\langle\boldsymbol{\lambda}_N|\Psi_0\rangle}{\sqrt{\langle\boldsymbol{\lambda}_N|\boldsymbol{\lambda}_N\rangle}}=\prod\limits_{j=1}^{N/2}\mathcal{F}(\lambda_j)\times\sqrt{\frac{\det G_{N/2}^+}{\det G_{N/2}^-}},
\end{align}
where $\mathcal{F}(\lambda)$ is a state-dependent function and $\det G_{N/2}^{\pm}$ are the Gaudin-like determinants.
We will prove that for the crosscap states the overlaps indeed take the form~\eqref{overlaps} with $\mathcal{F}(\lambda)=1$. We follow the method proposed in~\cite{Jiang:2020sdw} with important modifications.
\paragraph{Coordinate Bethe ansatz}
An $N$-particle eigenstate $\left|\boldsymbol{\lambda}_{N}\right\rangle$ for both the XXX and XXZ spin chains take the following form
\begin{align}
\left|\boldsymbol{\lambda}_{N}\right\rangle=\sum_{\left\{\boldsymbol{n}_{N}\right\}} \chi\left(\boldsymbol{n}_{N}, \boldsymbol{\lambda}_{N}\right)\left|n_{1}, n_{2}, \cdots, n_{N}\right\rangle,
\end{align}
where
\begin{align}
\label{eq:basisN}
\left|n_{1}, n_{2}, \ldots, n_{N}\right\rangle=S_{n_{1}}^{+} S_{n_{2}}^{+} \cdots S_{n_{N}}^{+}|\Omega\rangle.
\end{align}
The summation over $\left\{\boldsymbol{n}_{N}\right\}$ means summing over all the possible particle positions with the following constraint
\begin{align}
0\le n_1<n_2<...<n_{N}\le L-1.
\end{align}
The wave function $ \chi\left(\boldsymbol{n}_{N}, \boldsymbol{\lambda}_{N}\right)$ is given by
\begin{align}
\label{eq:BetheWaveFunc}
\chi\left(\boldsymbol{n}_{N}, \boldsymbol{\lambda}_{N}\right)=\sum_{\sigma \in S_{N}} \prod_{j>k} f\left(\lambda_{\sigma_{j}}-\lambda_{\sigma_{k}}\right) \prod_{j=1}^{N} e^{i p(\lambda_{\sigma_{j}}) n_{j}},
\end{align}
where the explicit form of $p(\lambda)$ and $f(\lambda)$ depend on the model. The rapidities $\boldsymbol{\lambda}_{N}$ satisfy the Bethe ansatz equations
\begin{align}
e^{i p\left(\lambda_{j}\right) L} \prod_{\substack{k=1 \\ k \neq j}}^{N} \frac{f(\lambda_{j}-\lambda_{k})}{f(\lambda_{k}-\lambda_{j})}=1,\quad j=1,2,...,N. 
\end{align}
For the XXX and XXZ spin chains, the two functions are given by
\begin{itemize}
\item XXX spin chain:
\begin{align}
e^{i p(\lambda)}=\frac{\lambda-i / 2}{\lambda+i / 2}, \quad f(\lambda)=\frac{\lambda+i}{\lambda}.
\end{align}
\item XXZ spin chain:
\begin{align}
e^{i p(\lambda)}=\frac{\sinh(\lambda-i\eta / 2)}{\sinh (\lambda+i\eta / 2)}, \quad f(\lambda)=\frac{\sinh (\lambda+ i\eta)}{\sinh (\lambda)}.
\end{align}
where $\eta$ is related to the anisotropy by $\Delta=\cosh\eta$.
\end{itemize}
The Bethe states are the eigenstate of the Hamiltonian
\begin{align}
H\left|\boldsymbol{\lambda}_{N}\right\rangle=E_{N}\left(\boldsymbol{\lambda}_{N}\right)\left|\boldsymbol{\lambda}_{N}\right\rangle.
\end{align}
For the XXX spin chain, the eigenvalue reads
\begin{align}
\quad E_{N}\left(\boldsymbol{\lambda}_{N}\right)=-\sum_{j=1}^{N}\frac{2}{\lambda_{j}^2+1/4}.
\end{align}
For the XXZ spin chain, the eigenvalue reads
\begin{align}
\quad E_{N}\left(\boldsymbol{\lambda}_{N}\right)=\sum_{j=1}^{N}\frac{4 \sinh ^{2} \eta}{\cos (2 \lambda_{j})-\cosh \eta}.
\end{align}
\paragraph{Norm of Bethe states}
The norm of the on-shell Bethe states takes the following form~\cite{Korepin:1982gg}
\begin{align}
\left\langle\boldsymbol{\lambda}_{N}|\boldsymbol{\lambda}_{N}\right\rangle=\prod_{j=1}^{N} \frac{1}{p^{\prime}\left(\lambda_{j}\right)} \prod_{j<k}^{N} f\left(\lambda_{j}-\lambda_{k}\right) f\left(\lambda_{k}-\lambda_{j}\right) \times \operatorname{det}G_{N},
\end{align}
where $G_{N}$ is the Gaudin matrix whose elements are given by
\begin{align}
&G_{j k}=\delta_{j k}\left(p^{\prime}\left(\lambda_{j}\right) L+\sum_{l=1}^{N} \varphi\left(\lambda_{j}-\lambda_{l}\right)\right)-\varphi\left(\lambda_{j}-\lambda_{k}\right)\,,\\\nonumber
&\varphi(\lambda)=-i\frac{\mathrm{d}}{\mathrm{d}\lambda}\log\left(\frac{f(\lambda)}{f(-\lambda)}\right).
\end{align}
If the Bethe roots are parity even~\eqref{eq:paired}, the norm further factorizes
\begin{align}
\langle\boldsymbol{\lambda}_{N}|\boldsymbol{\lambda}_{N}\rangle&=\prod_{j=1}^{N / 2} \frac{f\left(2 \lambda_{j}\right) f\left(-2 \lambda_{j}\right)}{\left(p^{\prime}\left(\lambda_{j}\right)\right)^{2}} \prod_{1 \leq j<k \leq N / 2}\left[\bar{f}\left(\lambda_{j}, \lambda_{k}\right)\right]^{2} \times \operatorname{det} G_{N/2}^{+} \operatorname{det} G_{N/2}^{-},
\end{align}
where 
\begin{align}
&G_{j k}^{\pm}=\delta_{j k}\left(p^{\prime}\left(\lambda_{j}\right) L+\sum_{l=1}^{N / 2} \varphi^{+}\left(\lambda_{j}, \lambda_{l}\right)\right)-\varphi^{\pm}\left(\lambda_{j}, \lambda_{k}\right),\\
&\varphi^{\pm}(\lambda, \mu)=\varphi(\lambda-\mu) \pm \varphi(\lambda+\mu),\\
&\bar{f}(\lambda, \mu)=f(\lambda-\mu) f(\lambda+\mu) f(-\lambda-\mu) f(-\lambda+\mu).
\end{align}
\paragraph{Some notations}
Following~\cite{Jiang:2020sdw}, it is convenient to introduce the notations
\begin{align}
\label{notation-1}
l_{j}=e^{i p\left(\lambda_{j}\right)},\quad f(l_{j},l_{k})=f(\lambda_{j}-\lambda_{k}).
\end{align} 
Then the wave function becomes 
\begin{align}
\chi\left(\boldsymbol{n}_{N}, \boldsymbol{\lambda}_{N}\right)=\sum_{\sigma \in S_{N}} \prod_{j>k} f\left(l_{\sigma_{j}},l_{\sigma_{k}}\right) \prod_{j=1}^{N} l_{\sigma_{j}}^{n_{j}},
\end{align}
and the Bethe equations can be written as
\begin{align}
a_{j}=l_j^{L}=\prod_{\substack{k=1 \\ k \neq j}}^{N} \frac{f\left(l_{k}, l_{j}\right)}{f\left(l_{j}, l_{k}\right)},\quad j=1,2,...,N.
\end{align}
\paragraph{Crosscap overlaps}
Now we turn to the overlaps. We first consider the overlaps between the crosscap state and the $N$-particle basis states \eqref{eq:basisN}. From the definition of the crosscap state~\eqref{Heisenberg-crosscap}, we find
\begin{align}
\label{eq:overlapCBasis}
\langle\mathcal{C}|n_1,n_2,...,n_{N}\rangle=&\prod_{i=1}^{N/2}\delta_{n_i,n_{i+N/2}-L/2},
\end{align}
which implies that the particles must appear in pairs with distance $L/2$. Therefore both $L$ and $N$ should be even. We can split the valid $N$-particle positions into two parts
\begin{align}
\{\boldsymbol{n}_{N}\}_C:=&\{\boldsymbol{n}_{N/2}\}\cup\{\boldsymbol{n}_{N/2}+\tfrac{L}{2}\},\\
\{\boldsymbol{n}_{N/2}\}:=&\{n_1,n_2,...,n_{N/2}| 0\le n_1<n_2<...<n_{N/2}\le \tfrac{L}{2}-1\},
\end{align} 
where the second part is completely determined by the first part. Using \eqref{eq:overlapCBasis}, the overlap of the crosscap state and a Bethe state reads
\begin{align}
\label{overlap-1}
\langle\mathcal{C}|\boldsymbol{\lambda}_{N}\rangle=\mathcal{S}_{N}(\boldsymbol{l}_{N},\boldsymbol{a}_{N})=\sum_{\sigma \in S_{N}} \prod_{j>k} f\left(l_{\sigma_{j}},l_{\sigma_{k}}\right)\sum_{\{\boldsymbol{n}_{N/2}\}} \prod_{j=1}^{N/2} l_{\sigma_{j}}^{n_{j}}l_{\sigma_{(j+N/2)}}^{n_{j}+L/2}.
\end{align}
We introduce the summation function
\begin{align}
\label{eq:defSummation}
B_{N}(\boldsymbol{l}_{N}|L)=\sum_{\{\boldsymbol{n}_{N/2}\}} \prod_{j=1}^{N/2} l_{j}^{n_{j}}l_{j+N/2}^{n_{j}+L/2},
\end{align}
for later convenience. To see how the method works, let us first consider the simplest 2-particle state.
\paragraph{$2$-particle states}  
The overlap can be calculated straightforwardly
\begin{align}
\label{eq:2particleOverlap}
\langle\mathcal{C}|\boldsymbol{\lambda}_{2}\rangle=
&f(l_2,l_1)\frac{l_2^{L/2}\left(1-(l_1l_2)^{L/2}\right)}{1-l_1l_2}+f(l_1,l_2)\frac{l_1^{L/2}\left(1-(l_1l_2)^{L/2}\right)}{1-l_1l_2}.
\end{align}
The Bethe equations read
\begin{align}
a_{1}=l_{1}^{L}=\frac{f(l_2,l_{1})}{f(l_1,l_2)},\qquad a_{2}=l_{2}^{L}=\frac{f(l_1,l_2)}{f(l_2,l_1)}.
\end{align}
For an on-shell Bethe state, the Bethe equation implies that $(l_1l_2)^{L/2}=1$. If we naively substituting this into \eqref{eq:2particleOverlap}, we find that the overlap is vanishing for any on-shell Bethe state. The key point here is to notice that we need to \emph{first} take the paired rapidities limit $l_1l_2\to 1$ and \emph{then} impose Bethe ansatz equations, which leads to
\begin{align}
\langle\mathcal{C}|\boldsymbol{\lambda}_{2}\rangle
=&L\sqrt{f(2\lambda_1)f(-2\lambda_1)}.
\end{align} 
The overlap can be written in Gaudin determinant form
\begin{align}
\label{G-initial-condition}
\langle\mathcal{C}|\boldsymbol{\lambda}_{2}\rangle=\frac{1}{p'(\lambda_1)}\sqrt{f(2\lambda_1)f(-2\lambda_1)}\times \operatorname{det} G_1^{+},\quad \operatorname{det} G_1^{+}=p'(\lambda_1)L.
\end{align}
From this simple example, we learned that the non-vanishing overlap is obtained by taking the paired rapidities limit before imposing Bethe ansatz equations.
\paragraph{$N$-particle states} 
For $N$-particle states, the summation function defined in \eqref{eq:defSummation} can be written as
\begin{align}
\label{summation-N-1}
B_{N}(\boldsymbol{l}_{N}|L)=&\prod_{j=1}^{N/2}l_{j+N/2}^{L/2}\sum_{n_1=0}^{\frac{L-N}{2}+1}\sum_{n_2=n_1+1}^{\frac{L-N}{2}+2}...\sum_{n_{N/2}=n_{N/2-1}+1}^{\frac{L}{2}}\prod_{j=1}^{N/2}\left(l_{j}l_{j+N/2}\right)^{n_{j}}.
\end{align}
If we ignore the overall factor in~\eqref{summation-N-1}, the summation function basically becomes a summation over half of the spin chain without constraints, except that we have $l_{j}l_{j+N/2}$ instead of $l_j$ in the summand. The latter summation function can be computed by using a recursion relation~\cite{Jiang:2020sdw,deLeeuw:2015hxa}. This allows us to obtain an explicit albeit slightly involved expression for $B_{N}(\boldsymbol{l}_{N}|L)$
\begin{align}
B_{N}(\boldsymbol{l}_{N}|L)=&\sum_{j=0}^{N/2} B_{N,j}(\boldsymbol{l}_{N}|L),
\end{align}
where
\begin{align}
 B_{N,j}(\boldsymbol{l}_{N}|L)=\frac{(-1)^{j}\prod_{k=j+1}^{j+N/2}\left(a_{k}\right)^{1/2}\prod_{k=j+N/2+1}^{N} a_{k} \prod_{k=2}^{j}(l_{k}l_{k+N/2})^{k-1}}{\prod_{k=j+1}^{N/2}\left(\prod_{i=j+1}^{k} l_{i}l_{i+N/2}-1\right)\prod_{k=1}^{j}\left(\prod_{i=k}^{j} l_{i}l_{i+N/2}-1\right)}.
\end{align}
The summation function is a rational function of $l_{j}$.
In order to take the paired rapidity limit, we first consider the behavior of $B_{N}(\boldsymbol{l}_{N}|L)$ near the pole at $l_ml_{m+N/2}=1$. There are two terms $B_{N,m-1}(\boldsymbol{l}_{N}|L)$ and $B_{N,m}(\boldsymbol{l}_{N}|L)$ that contain this pole. Taking the sum of these two terms and using $l_ml_{m+N/2}=1$ for the regular part, we obtain
\begin{align}
&B_{N,m-1}(\boldsymbol{l}_{N}|L)+B_{N,m}(\boldsymbol{l}_{N}|L)\nonumber\\
=&\frac{\left[\left(a_{m}a_{m+N/2}\right)^{1/2}-1\right]\left(a_{m+N/2}\right)^{1/2}}{l_{m}l_{m+N/2}-1}\times\nonumber\\\quad &\frac{(-1)^{m-1}\prod_{k=m+1}^{m+N/2-1}\left(a_{k}\right)^{1/2}\prod_{k=m+N/2+1}^{N} a_{k} \prod_{k=2}^{m-1}(l_{k}l_{k+N/2})^{k-1}}{\prod_{k=m+1}^{N/2}\left(\prod_{i=m+1}^{k} l_{i}l_{i+N/2}-1\right)\prod_{k=1}^{m-1}\left(\prod_{i=k}^{m-1} l_{i}l_{i+N/2}-1\right)},
\end{align}
Notice that the second line is nothing but $B_{N,m}$ with two particles at $m$ and $m+N/2$ removed. Therefore near the pole $l_ml_{m+N/2}=1$, we have
\begin{align}
\label{eq:mainrecursion}
B_{N}(\boldsymbol{l}_{N}|L)\sim&\frac{\left(a_{m}a_{m+N/2}\right)^{1/2}-1}{l_{m}l_{m+N/2}-1}\left(a_{m+N/2}\right)^{1/2}B_{N-2,m-1}(\{1,...\cancel{m}...\cancel{m+N/2}...,N\}|L).
\end{align}
Plugging back to the overlap formula~\eqref{overlap-1}, we also need to multiply a factor in front of $B_{N}(\boldsymbol{l}_{N}|L)$ and then sum over all the permutations. Since we focus on the pole $l_{m}l_{m+N/2}=1$, we also need to separate out the $l_{m}$ and $l_{m+N/2}$ dependent part for the multiplying factor. Note that the exchange of $l_{m}$ and $l_{m+N/2}$ preserve the relative position of $m$ and $m+L/2$, which gives the same pole. After factorizing the $l_{m}l_{m+N/2}=1$ pole, we find
\begin{align}
\label{S-recursion}
&\mathcal{S}_{N}(\boldsymbol{l}_{N},\boldsymbol{a}_{N})=\sum_{\sigma \in S_{N}} \prod_{j>k} f\left(l_{\sigma_{j}},l_{\sigma_{k}}\right)B_{N}(\sigma\boldsymbol{l}_{N}|L)\nonumber\\
\sim&\frac{\left(a_{m}a_{m+N/2}\right)^{1/2}-1}{l_{m}l_{m+N/2}-1}\left(F_{m}+F_{m+N/2}\right)\nonumber\\
&\times\sum_{\sigma \in S_{N-2}} \prod_{\substack{j>k \\ j,k \neq m,m+N/2}} f\left(l_{\sigma_{j}},l_{\sigma_{k}}\right)B_{N-2,m-1}(\sigma\{1,...\cancel{m}...\cancel{m+N/2}...,N\}|L),
\end{align}
where the last line does not depend on $l_{m}$ and $l_{m+N/2}$. 
The sum of $F_{m}$ and $F_{m+N/2}$ reads
\begin{align}
\label{factor}
F_{m}+F_{m+N/2}=&\prod_{j=m+1}^{m+N/2-1}\left[\frac{f\left(l_{j},l_{m}\right) f\left(l_{j},l_{m+N/2}\right)}{f\left(l_{m},l_{j}\right) f\left(l_{m+N/2},l_{j}\right)}\right]^{1/2}\prod_{j=m+N/2+1}^{N} \frac{f\left(l_{j},l_{m}\right) f\left(l_{j},l_{m+N/2}\right)}{f\left(l_{m},l_{j}\right) f\left(l_{m+N/2},l_{j}\right)}\nonumber\\
&\times\prod_{\substack{j=1 \\ j \neq m,m+N/2}}^{N} f\left(l_{m},l_{j}\right) f\left(l_{m+N/2},l_{j}\right)\times \mathbf{F}(l_{m},l_{m+N/2}),
\end{align}
where
\begin{align}
\label{F-factor}
\mathbf{F}(l_{m},l_{m+N/2})=&\prod_{j=m+1}^{m+N/2-1}\left[\frac{f\left(l_{j},l_{m}\right)f\left(l_{m+N/2},l_{j}\right)}{ f\left(l_{m},l_{j}\right)f\left(l_{j},l_{m+N/2}\right)}\right]^{1/2}\times f(l_{m+N/2},l_m)\left(a_{m+N/2}\right)^{1/2}\nonumber\\
+&\prod_{j=m+1}^{m+N/2-1}\left[\frac{f\left(l_{j},l_{m+N/2}\right)f\left(l_{m},l_{j}\right)}{ f\left(l_{m+N/2},l_{j}\right)f\left(l_{j},l_{m}\right)}\right]^{1/2}\times f(l_{m},l_{m+N/2})\left(a_{m}\right)^{1/2}.
\end{align}
The main new feature of the crosscap state is that the function $\mathbf{F}(l_{m},l_{m+N/2})$ depends on not only $l_{m},l_{m+N/2}$ but also $l_j$ for $m<j<m+N/2$, while for the integrable boundary states considered in~\cite{Jiang:2020sdw}, the poles appear at neighboring $l_ml_{m+1}=1$ and the function $\mathbf{F}$ just depends on $l_m$ and $l_{m+1}$. 
We introduce the the modified parameters
\begin{align}
a^{\text{mod}}_{j}=&\frac{f\left(l_{j},l_{m}\right) f\left(l_{j},l_{m+N/2}\right)}{f\left(l_{m},l_{j}\right) f\left(l_{m+N/2},l_{j}\right)}a_{j},\quad 1\le j\le N\,,
\end{align}
so that the first line on the right hand side in~\eqref{factor} can be absorbed in $B_{N-1,m-1}$ by making the replacement $a_{j}\to a^{\text{mod}}_{j}$.  Then the recursion relation~\eqref{S-recursion} can be written as
\begin{align}
\label{eq:recurseS}
\mathcal{S}_{N}(\boldsymbol{l}_{N}|L)\sim&\frac{\left(a_{m}a_{m+N/2}\right)^{1/2}-1}{l_{m}l_{m+N/2}-1}\mathbf{F}(l_{m},l_{m+N/2})\prod_{\substack{j=1 \\ j \neq m}}^{N/2} \bar{f}\left(l_{m},l_{j}\right)\nonumber\\
&\times\mathcal{S}^{\text{mod}}_{N-2}(\{1,...\cancel{m}...\cancel{m+N/2}...,N\}|L)
\end{align}
Now we consider the paired rapidity limit
\begin{align}
\lambda_{j+N/2}\to -\lambda_{j},\qquad j=1,2,...,N/2.
\end{align}
In this limit, $\mathbf{F}(l_m,l_{m+N/2})$ simplifies drastically 
\begin{align}
\label{limit-F}
&\mathbf{F}(l_{m},l_{m+N/2})\to 2\sqrt{f(-2\lambda_{m})f(2\lambda_{m})}\,.
\end{align} 
The details can be found in appendix~\ref{app:A}. Let us denote the paired rapidity limit of $\mathcal{S}_{N}(\boldsymbol{l}_{N}|L)$  by $D\left(\boldsymbol{\lambda}_{N / 2}, \boldsymbol{m}_{N / 2}|L\right)$, in which we have introduced the new parameter
\begin{align}
m_{j}=-i \frac{\mathrm{d}}{\mathrm{d}\lambda_{j}} \log (a_{j})=p'(\lambda_{j})L.
\end{align}
The recursion relation \eqref{eq:recurseS} implies that 
\begin{align}
\label{diff-eq-D}
\dfrac{\partial D(\boldsymbol{\lambda}_{N/2},\boldsymbol{m}_{N/2}|L)}{\partial m_{m}}=\frac{\sqrt{f(-2\lambda_{m})f(2\lambda_{m})}}{p'(\lambda_{m})}\prod_{\substack{j=1 \\ j \neq m}}^{N/2}\bar{f}(\lambda_{m},\lambda_{j})\times D(\boldsymbol{\lambda}_{N/2-1},\boldsymbol{m}_{N/2-1}^{\text{mod}}|L),
\end{align}
where the modified parameter is given by
\begin{align}
\label{mod-relation-m}
m^{\text{mod}}_{j}=-i \frac{\mathrm{d}}{\mathrm{d}{\lambda}_{j}} \log \left(a^{\text{mod}}_{j}\right)=m_j+\varphi^{+}(\lambda_{j},\lambda_{m}).
\end{align}
To write the recursion relation in a nicer form, let us define $\tilde{D}\left(\boldsymbol{\lambda}_{N / 2}, \boldsymbol{m}_{N / 2}|L\right)$ via
\begin{align}
D\left(\boldsymbol{\lambda}_{N / 2}, \boldsymbol{m}_{N / 2}|L\right)=\prod_{j=1}^{N / 2} \frac{\sqrt{f(-2\lambda_{j})f(2\lambda_{j})}}{p^{\prime}\left(\lambda_{j}\right)} \prod_{1 \leq j<k \leq N / 2} \bar{f}\left(\lambda_{j}, \lambda_{k}\right) \tilde{D}\left(\boldsymbol{\lambda}_{N / 2}, \boldsymbol{m}_{N / 2}|L\right).
\end{align}
It follows from~\eqref{diff-eq-D} that the new function satisfies
\begin{align}
\label{eq:recurseDtilde}
\frac{\partial \tilde{D}\left(\boldsymbol{\lambda}_{N / 2}, \boldsymbol{m}_{N / 2} |L\right)}{\partial m_{m}}=\tilde{D}\left(\boldsymbol{\lambda}_{N / 2-1}, \boldsymbol{m}_{N / 2-1}^{\text{mod}}|L\right),
\end{align}
where it is understood that $m_{m}$ is not included on the right hand side. The initial condition for this differential equation is given in~\eqref{G-initial-condition}. Following the same arguments as in~\cite{Korepin:1982gg,Jiang:2020sdw}, the unique solution to \eqref{eq:recurseDtilde} is nothing but the Gaudin determinant
\begin{align}
\tilde{D}\left(\boldsymbol{\lambda}_{N / 2}, \boldsymbol{m}_{N / 2}| L\right)=\det G_{N/2}^{+}.
\end{align}
Finally, we obtain the exact on-shell overlaps of crosscap states and Bethe states
\begin{align}
\langle\mathcal{C}|\boldsymbol{\lambda}_{N}\rangle&=\prod_{j=1}^{N / 2} \frac{\sqrt{f\left(2 \lambda_{j}\right) f\left(-2 \lambda_{j}\right)}}{p^{\prime}\left(\lambda_{j}\right)} \prod_{1 \leq j<k \leq N / 2}\bar{f}\left(\lambda_{j}, \lambda_{k}\right) \times \operatorname{det} G_{N/2}^{+},
\end{align}
which leads to 
\begin{align}
\label{normalize-overlap-0}
\frac{\langle\mathcal{C}|\boldsymbol{\lambda}_{N}\rangle}{\sqrt{\langle\boldsymbol{\lambda}_{N}|\boldsymbol{\lambda}_{N}\rangle}}=\sqrt{\frac{\det G_{N/2}^{+}}{\det G_{N/2}^{-}}}.
\end{align} 
Therefore we indeed find a trivial prefactor. This fact seems to be universal for crosscap states defined in all integrable models so far. From our derivations, it is clear that this is the case for both XXX and XXZ spin chains. The exact crosscap overlap formula in $\mathfrak{gl}(N)$ symmetric spin chains also takes the same form~\cite{Gombor:2022deb}. We will see that it is also the case for the Lieb-Liniger model.
\subsection{Overlap formula for the non-compact chain}
\label{sec:2.3}
In this subsection, we generalize the above consideration to the non-compact $SL(2,R)$ spin chain. The Hilbert space is spanned by the following $N$-particle basis
\begin{align}
\left|n_{1}, n_{2}, \cdots, n_{N}\right\rangle \equiv S_{n_1}^{+}S_{n_2}^{+}\cdots S_{n_{N}}^{+}|\Omega\rangle,
\end{align}
where $|\Omega\rangle$ denotes the pseudovacuum. Since for the non-compact chain we can excite multiple particles on one site, when we take the sum over the magnon positions we have   
\begin{align}
\label{eq:restrictn}
0\le n_1\le n_2\le...\le n_{N}\le L-1.
\end{align}
This model is also solvable by coordinate Bethe ansatz. The Bethe states take the same form as for the compact chains, except the restriction for the summation over $n_j$ is now given by \eqref{eq:restrictn}, and the two functions in the Bethe wavefunction reads
\begin{align}
e^{i p(\lambda)}=\frac{\lambda-i / 2}{\lambda+i / 2}, \qquad f(\lambda)=\frac{\lambda-i}{\lambda}.
\end{align}
We shall follow the same procedure as before. From the definition of the crosscap state~\eqref{SL-crosscap}, for basis states satisfying \eqref{eq:restrictn} we have
\begin{align}
\label{eq:overlapCCnoncompact}
\langle\mathcal{C}|n_1,n_2,...,n_{N}\rangle=\prod_{j=1}^{N/2}\delta_{n_j,n_{j+N/2}-L/2}\,.
\end{align}
We can again divide the position of $N$-particle states into two sets 
\begin{align}
\{\boldsymbol{n}_{N}\}_{C}:=&\{\boldsymbol{n}_{N/2}\}\cup\{\boldsymbol{n}_{N/2}+\tfrac{L}{2}\},\\
\{\boldsymbol{n}_{N/2}\}:=&\{n_1,n_2,...,n_{N/2}| 0\le n_1\le n_2\le...\le n_{N/2}\le \tfrac{L}{2}-1\}.
\end{align} 
Using \eqref{eq:overlapCCnoncompact}, the overlap reads
\begin{align}
\label{overlap-2}
\langle\mathcal{C}|\boldsymbol{\lambda}_{N}\rangle=\sum_{\sigma \in S_{N}} \prod_{j>k} f\left(l_{\sigma_{j}},l_{\sigma_{k}}\right)\sum_{\{\boldsymbol{n}_{N/2}\}} \prod_{j=1}^{N/2} l_{\sigma_{j}}^{n_{j}}l_{\sigma_{(j+N/2)}}^{n_{j}+L/2}.
\end{align}
Similarly, it is convenient to introduce the summation function
\begin{align}
\label{summation-N-2}
B_{N}(\boldsymbol{l}_{N}|L)
=&\prod_{j=1}^{N/2}l_{j+N/2}^{L/2}\sum_{n_1=0}^{L/2-1}\sum_{n_2=0}^{L/2-1}...\sum_{n_{N/2}=0}^{L/2-1}\prod_{j=1}^{N/2}\left(l_{j}l_{j+N/2}\right)^{n_{j}}.
\end{align}
For the non-compact chain, $B_{N}$ can also be computed by a recursion relation, leading to
\begin{align}
B_{N}(\boldsymbol{l}_{N}|L)=&\sum_{j=0}^{N/2} B_{N,j}(\boldsymbol{l}_{N}|L),
\end{align}
where 
\begin{align}
B_{N,j}(L)=\frac{(-1)^{j}\prod_{k=j+1}^{j+N/2}\left(a_{k}\right)^{1/2}\prod_{k=j+N/2+1}^{N} a_{k} \prod_{k=j+1}^{N/2}(l_{k}l_{k+N/2})^{N/2-k}}{\prod_{k=j+1}^{N/2}\left(\prod_{i=j+1}^{k} l_{i}l_{i+N/2}-1\right)\prod_{k=1}^{j}\left(\prod_{i=k}^{j} l_{i}l_{i+N/2}-1\right)}.
\end{align}
The main observation is that, although the summation functions are different, their behavior near the pole $l_ml_{m+N/2}$ are the same
\begin{align}
B_{N}(\boldsymbol{l}_{N}|L)\sim&\frac{\left(a_{m}a_{m+N/2}\right)^{1/2}-1}{l_{m}l_{m+N/2}-1}\left(a_{m+N/2}\right)^{1/2}B_{N-2,m-1}(\{1,...\cancel{m}...\cancel{m+N/2}...,N\}|L).
\end{align}
This leads to the same $\mathbf{F}(l_{m},l_{m+N/2})$ function and the same recursion relation for the overlap. As a consequence, we get the crosscap overlap formula 
\begin{align}
\label{normalize-overlap-2}
\frac{\langle\mathcal{C}|\boldsymbol{\lambda}_{N}\rangle}{\sqrt{\langle\boldsymbol{\lambda}_{N}|\boldsymbol{\lambda}_{N}\rangle}}=\sqrt{\frac{\det G_{N/2}^{+}}{\det G_{N/2}^{-}}}.
\end{align} 
\par
Before ending the section, let us comment on the universality of the crosscap overlaps. In fact, the key point is the overlaps between crosscap states and $N$-particle basis gives a product of the Kronecker deltas, which imposes strong constraints on the particle positions. Such constraints reflects the geometric origin of the crosscap state. It effectively reduces the Bethe wave function of $N$ particles to be the one with $N/2$ particles, except for the replacements $l_{j}\to l_{j}l_{j+N/2}$, see~\eqref{summation-N-1} and~\eqref{summation-N-2}. The overlaps are the sum of Bethe wave functions which satisfy the constraints on the particle positions. For the crosscap constaints, the overlaps turn to be the sum of reduced Bethe wave function of $N/2$ particles without constraints on the particle positions. After taking the paired rapidities limit, the sum of the reduced Bethe wave functions is actually the norm of the Bethe states except for the replacement of Gaudin-like determinant $\det G_{N/2}^{-}\to \det G_{N/2}^{+}$, which hence leads to the trivial prefactor for the overlaps.
\section{Crosscap state of Lieb-Liniger model}
\label{sec:3}
In this section, we present the derivations of our proposal for the crosscap state of the Lieb-Liniger model \eqref{eq:CCstart}. We will then prove its integrability and derive the exact overlap formula.
\subsection{Crosscap state in Lieb-Liniger model}
\label{sec:3.1}
In the second quantized form, the Hamiltonian of the Lieb-Liniger model is given by
\begin{align}
H=\int_0^{\ell}\mathrm{d}x\left[\partial_x\Phi^\dagger(x)\partial_x\Phi(x)+c\,\Phi^\dagger(x)\Phi^\dagger(x)\Phi(x)\Phi(x)\right],
\end{align}
where the bosonic fields satisfy the usual commutation relations
\begin{align}
[\Phi(x),\Phi^{\dagger}(y)]=\delta(x-y),\quad [\Phi(x),\Phi(y)]=[\Phi^{\dagger}(x),\Phi^{\dagger}(y)]=0.
\end{align}
We consider the periodic boundary condition with system size $\ell$. This model can be solved by coordinate Bethe ansatz as well as the Quantum Inverse Scattering Method (QISM)~\cite{1993Quantum}. We shall make use of both approaches in what follows. For the proof of integrability, it is more convenient to use QISM, but it is necessary to first define the model on the lattice and then take the continuum limit. For the derivation of exact overlap formula, we make use of coordinate Bethe ansatz.\par

As mentioned before, Lieb-Liniger model can be obtained by taking continuum limit of integrable lattice models. There are at least two ways to achieve this. The first one is by taking the special continuum limit of the XXZ spin chain after performing the Dyson-Maleev transformation~\cite{Golzer_1987}. The second one is taking the continuum limit of a generalized XXX spin chain~\cite{Izergin:1982ry}. Our strategy is staring from the crosscap state in spin chains, and then taking the continuum limit. We consider both approaches, and it turns out that they lead to the same crosscap state in the Lieb-Liniger model. 
\paragraph{Method I: Dyson-Maleev transformation} The Dyson-Maleev transformation maps the local spin operators to the bosonic operators
\begin{align}
S^{+}_{i}=a_{i}^{\dagger}(1-a_{i}^{\dagger}a_{i}),\quad S^{-}_{i}=a_{i},\quad S^{z}_{i}=-\frac{1}{2}+a_{i}^{\dagger}a_{i}\,,
\end{align}
where the bosonic operators satisfy the canonical commutation relations
\begin{align}
[a_{i},a^{\dagger}_{j}]=\delta_{ij},\quad [a_{i},a_{j}]=[a^{\dagger}_{i},a^{\dagger}_{j}]=0.
\end{align}
Applying the transformation to the crosscap state of the XXZ spin chain~\eqref{Heisenberg-crosscap}, we obtain
\begin{align}
\label{DM-crosscap}
|\mathcal{C}\rangle=&\prod_{j=1}^{L/2}\left(1+a^{\dagger}_{j}a^{\dagger}_{j+L/2}\right)|\Omega\rangle,
\end{align}
where we have used the fact
\begin{align}
a_{i}|\Omega\rangle=S^{-}_{i}|\Omega\rangle=0.
\end{align}  
\par
Let us consider a XXZ spin chain of length $L$. We denote the lattice spacing by $\delta$. The system size of the continuum model is $\ell=L\delta$. The bosonic operators can be written in terms of Fourier modes
\begin{align}
a_{n}=\frac{1}{\sqrt{L}}\sum_{k}e^{-ik x_n}\tilde{a}_{k},\quad x_n=n\delta.
\end{align}
In this representation, the continuum limit is obtained by
\begin{align}
\sum_{k}\to\frac{\ell}{2\pi}\int\mathrm{d}k,\qquad \tilde{a}_{k}\to \left(\frac{2\pi}{\ell}\right)^{1/2}\tilde{\Phi}_{k}.
\end{align}
Then one transforms back to real space by the inverse Fourier transformation
\begin{align}
\tilde{\Phi}_{k}=\frac{1}{\sqrt{2\pi}}\int\mathrm{d}x\,e^{ikx}\Phi(x),\quad k=\frac{2\pi m}{\ell}.
\end{align}
Applying above procedure to~\eqref{DM-crosscap}, we arrive at the crosscap state
\begin{align}
\label{crosscap-0}
|\mathcal{C}\rangle=&\exp\left(\int_0^{\ell/2}dx{\Phi}^{\dagger}(x){\Phi}^{\dagger}(x+\ell/2)\right)|\Omega\rangle,
\end{align}
where we assume the pseudovacuum in spin chain corresponds to the Fock vacuum $|\Omega\rangle$ of the Lieb-Liniger model in the continuum limit. Details of the derivation can be found in appendix~\ref{app:B}.

\paragraph{Method II: Generalized XXX model}
In this approach, we first discretize the Lieb-Liniger model by picking $L$ points on the interval $[0,\ell]$ located at $x_{n}=\Delta n,x_{L}=\ell$~\cite{Izergin:1982ry}. We then define the operators
\begin{align}
\psi_{n}=\frac{1}{\sqrt{\Delta}}\int_{x_{n-1}}^{x_n}\Phi(x)\,\mathrm{d}x,\quad \psi^{\dagger}_{n}=\frac{1}{\sqrt{\Delta}}\int_{x_{n-1}}^{x_n}\Phi^{\dagger}(x)\,\mathrm{d}x.
\end{align}
One can  check the operators satisfy
\begin{align}
[\psi_{n},\psi^{\dagger}_{m}]=\delta_{mn},\quad [\psi_{n},\psi_{m}]=[\psi^{\dagger}_{n},\psi^{\dagger}_{m}]=0.
\end{align}
In the lattice model, the pseudovacuum is identified with the Fock vacuum which satisfies 
\begin{align}
\Phi(x)|0\rangle=\psi_{n}|0\rangle=0.
\end{align}
The quantum Lax operator takes the form
\begin{align}
L_{n}(u)=\left(\begin{array}{cc}
1-\frac{i u \Delta}{2}+\frac{c\Delta}{2} \psi_{n}^{\dagger} \psi_{n} & -i\sqrt{c\Delta}\psi_{n}^{\dagger} \rho_{n}^{+}\\
i \sqrt{c\Delta}\rho_{n}^{-} \psi_{n}   & 1+\frac{i u \Delta}{2}+\frac{c\Delta}{2} \psi_{n}^{\dagger} \psi_{n}
\end{array}\right),
\end{align}
where the operator $\rho^{\pm}_{n}$ satisfy two constraints:
\begin{align}
\rho_{n}^{\pm}=\rho_{n}^{\pm}\left(\psi_{n}^{\dagger} \psi_{n}\right),\quad \rho_{n}^{+} \rho_{n}^{-}=1+\frac{c\Delta}{4} \psi_{n}^{\dagger} \psi_{n}.
\end{align}
For example, we can take
\begin{align}
\rho_{n}^{-}=1, \qquad \rho_{n}^{+}=1+\frac{c\Delta}{4} \psi_{n}^{\dagger} \psi_{n}.
\end{align}
The transfer matrix is defined as usual
\begin{align}
\label{discrete-LL-transfer}
T(u)=\text{tr}\Big({L}_{1}(u){L}_{2}(u)\ldots{L}_{L}(u)\Big).
\end{align}
In order to define the crosscap state for the discretized Lieb-Liniger model, we make use of the fact that it is closely related to the \emph{generalized} XXX spin chain~\cite{Izergin:1982ry,Izergin:2009yc}. This can be seen easily by the following transformation of the Lax operator
\begin{align}
\tilde{L}_{n}(u)=&\sigma_{3}\sigma_{2}L_{n}(u)\sigma_{2}
=\frac{i\Delta}{2}u\textbf{1}\otimes\textbf{1}+\frac{c\Delta}{2}\left[\tilde{S}_n^{z}\otimes\sigma^3-\left(\tilde{S}_n^{+}\otimes\sigma^{-}+\tilde{S}_n^{-}\otimes\sigma^{+}\right)\right],
\end{align}
where we have introduced 
\begin{align}
\label{Definition-S}
\tilde{S}_n^{z}=&\frac{2}{c\Delta}+\psi_{n}^{\dagger} \psi_{n},\quad 
\tilde{S}_n^{-}=\frac{2i}{\sqrt{c\Delta}}\rho_{n}^{-} \psi_{n},\quad
\tilde{S}_n^{+}=\frac{2i}{\sqrt{c\Delta}}\psi_{n}^{\dagger} \rho_{n}^{+}.
\end{align}
One can verify that they satisfy the standard SU(2) algebra
\begin{align}
\left[\tilde{S}_{n}^{z}, \tilde{S}_{n}^{\pm}\right]=\pm \tilde{S}_{n}^{\pm}, \quad\left[\tilde{S}_{n}^{+}, \tilde{S}_{n}^{-}\right]=2 \tilde{S}_{n}^{z}.
\end{align}
The local spin operators act on the vacuum as following
\begin{align}
\tilde{S}_{n}^{-}|\Omega\rangle=0,\quad \tilde{S}_{n}^{+}|\Omega\rangle=\psi_{n}^{\dagger}|\Omega\rangle,\quad \tilde{S}_{n}^{z}|\Omega\rangle=\frac{2}{c\Delta}|\Omega\rangle.
\end{align}
For the generalized XXX spin chain, we can define the crosscap state using the local spin operators $\tilde{S}$
\begin{align}
|\mathcal{C}\rangle=\prod_{n=1}^{L/2}\left(1+\tilde{S}_{n}^{+}\tilde{S}_{n+\frac{L}{2}}^{+}\right)|\Omega\rangle.
\end{align} 
We then show that the crosscap state so defined is integrable in the sense of \eqref{Integrable-condition}. Firstly, one can verify
\begin{align}
\sigma_2\tilde{L}_{n}(u)\sigma_2|c\rangle\!\rangle_{n}=-\tilde{L}_{n+L/2}(-u)|c\rangle\!\rangle_{n}\,.
\end{align}
It follows that
\begin{align}
\sigma_2L_{n}(u)\sigma_2|c\rangle\!\rangle_{n}=L_{n+L/2}(-u)|c\rangle\!\rangle_{n}.
\end{align}
Then we have 
\begin{align}
T(u)|\mathcal{C}\rangle
=&\langle\mathcal{C}|\text{tr}\left[\Big(L_{1}(u)...L_{L/2}(u)\Big)\Big(L_{L/2+1}(u)...L_{L}(u)\Big)\right]\nonumber\\
=&\langle\mathcal{C}|\text{tr}\left[\Big(L_{1}(u)...L_{L/2}(u)\Big)\Big(\sigma_2L_{1}(-u)...L_{L/2}(-u)\sigma_2\Big)\right]\nonumber\\
=&\langle\mathcal{C}|\text{tr}\left[\Big(L_{1}(-u)...L_{L/2}(-u)\Big)\Big(\sigma_2L_{1}(u)...L_{L/2}(u)\sigma_2\Big)\right]\nonumber\\
=&\langle\mathcal{C}|\text{tr}\left[\Big(L_{1}(-u)...L_{L/2}(-u)\Big)\Big(L_{L/2+1}(-u)...L_{L}(-u)\Big)\right]\nonumber\\
=&T(-u)|\mathcal{C}\rangle.
\end{align}
In the continuum limit, we expect the crosscap state becomes an integrable boundary state for the Lieb-Liniger model. The continuum version of the crosscap states can be obtained
\begin{align}
|\mathcal{C}\rangle \equiv&\prod_{n=1}^{L/2}\left(1+\tilde{S}_{n}^{+}\tilde{S}_{n+\frac{L}{2}}^{+}\right)|\Omega\rangle\to\exp\left(\int_{0}^{\ell/2}\mathrm{d}x\,\Phi^{\dagger}(x)\Phi^{\dagger}(x+\ell/2)\right)|\Omega\rangle.
\end{align} 
Details about taking the continuum limit can be found in appendix~\ref{app:B}. The resulting state is indeed the same as the result from the first method. 
\par
Two comments about the crosscap state are in order. 
\begin{itemize}
\item By expanding the exponential function, we can write the crosscap state as
\begin{align}
\label{eq:expandC1}
|\mathcal{C}\rangle=\sum_{N=0}^{\infty}|\mathcal{C}_{2N}\rangle,
\end{align}
where the $2N$-particle state is given by 
\begin{align}
\label{eq:expandC2}
|\mathcal{C}_{2N}\rangle =\int_{T}\mathrm{d}^{N}x \prod_{j=1}^{N}\Phi^{\dagger}(x_{j}) \Phi^{\dagger}(x_{j}+\ell/2) |\Omega\rangle.
\end{align}
The factor $1/N!$ from the exponential is dropped because we have fixed the order of particle position $T:0\leq x_1< x_2<...<x_{N}\leq \ell/2$. The crosscap state is a superposition state of even number of particles. Therefore, the crosscap state can have non-vanishing overlaps with Bethe states with any even number of particles.
\item The crosscap state for Lieb-Liniger model can also be written in momentum space by performing a Fourier transformation
\begin{align}
\label{crosscap-1}
|\mathcal{C}\rangle=&\exp \left(\frac{i}{2 \pi} \sum_{m, n} K_{m n} \xi_{m}^{\dagger} \xi_{n}^{\dagger}\right)|\Omega\rangle, \quad K_{m n}=\frac{(-1)^{m}-(-1)^{n}}{m+n},
\end{align}
where we used
\begin{align}
\Phi(x)=\frac{1}{\sqrt{\ell}} \sum_{q} e^{i q x} \xi_{q}, \qquad q=\frac{2 \pi m}{\ell}.
\end{align}
This formula is similar to the boundary state in integrable quantum field theory, and the coefficient $K_{mn}$ plays the role of two-particle boundary amplitudes~\cite{Ghoshal:1993tm}.
\end{itemize}
\subsection{Exact overlap formula}
\label{sec:3.2}
In this subsection, we will derive the overlap of the crosscap state and an on-shell Bethe state using the coordinate Bethe ansatz. This method has been applied in the calculating the overlap of the BEC state and Bethe state in Lieb-Liniger model~\cite{Chen:2020xel}.\par

Using coordinate Bethe ansatz, the eigenstate is given by
\begin{align}
|\boldsymbol{\lambda}_{N}\rangle=
\frac{1}{\sqrt{N !}} \int_{0}^{\ell}\mathrm{d}^{N}x\, \chi_{N}\left(\boldsymbol{x}_{N}| \boldsymbol{\lambda}_{N}\right) \Phi^{\dagger}\left(x_{1}\right) \ldots \Phi^{\dagger}\left(x_{N}\right)|\Omega\rangle,
\end{align}
where $|\Omega\rangle$ is the Fock vacuum of the Lieb-Liniger model. The wave function is
\begin{align}
\chi_{N}\left(\boldsymbol{x}_{N}| \boldsymbol{\lambda}_{N}\right)=&\frac{1}{\sqrt{N !}}\sum_{\sigma\in S_{N}}\prod_{j>k}\left[\frac{\lambda_{\sigma_j}-\lambda_{\sigma_k}-i c \epsilon\left(x_{j}-x_{k}\right)}{\lambda_{\sigma_j}-\lambda_{\sigma_k}}\right]\exp \left(i \sum_{n= 1}^{N} x_{n} \lambda_{\sigma_n}\right),
\end{align}
where $\epsilon$ is the sign function. We consider the configuration space
\begin{align}
\label{domain}
T: 0\le x_1<x_2<...<x_N\le \ell\,.
\end{align}
Introducing
\begin{align}
f(\lambda)=\frac{\lambda-ic}{\lambda},\qquad l_{j}=e^{i\lambda_{j}}\,,
\end{align}
we can write the wave function as
\begin{align}
\chi_{N}\left(\boldsymbol{x}_{N}| \boldsymbol{\lambda}_{N}\right)=&\frac{1}{\sqrt{N !}}\sum_{\sigma\in S_{N}}\prod_{j>k}f(\lambda_{\sigma_j}-\lambda_{\sigma_k})\prod_{n=1}^{N}l_{\sigma_{n}}^{x_{n}}.
\end{align}
Similar to the spin chain, the norm of an on-shell Bethe state is given by the Gaudin determinant
\begin{align}
\left\langle\boldsymbol{\lambda}_{N}|\boldsymbol{\lambda}_{N}\right\rangle= \prod_{j > k} f\left(\lambda_{j}-\lambda_{k}\right)f\left(\lambda_{k}-\lambda_{j}\right) \operatorname{det} G_{N},
\end{align}
where the Gaudin matrix elements are
\begin{align}
G_{j k}=\delta_{j k}\left(\ell+\sum_{l=1}^{N} \varphi\left(\lambda_{j}-\lambda_{l}\right)\right)-\varphi\left(\lambda_{j}-\lambda_{k}\right),
\end{align}
with
\begin{align}
\label{LL-phi}
\varphi(\lambda)=\frac{2c}{\lambda^2+c^2}.
\end{align}
If the rapidities are paired, the norm factorizes
\begin{align}
\label{Bethe-norm-LL}
\left\langle\boldsymbol{\lambda}_{N}|\boldsymbol{\lambda}_{N}\right\rangle=\prod_{j=1}^{N / 2} f\left(2\lambda_{j}\right) f\left(-2\lambda_{j}\right) \prod_{1 \leq j<k \leq N / 2}\left[\bar{f}\left(\lambda_{j}, \lambda_{k}\right)\right]^{2} \operatorname{det} G_{N/2}^{+} \operatorname{det} G_{N/2}^{-},
\end{align}
where
\begin{align}
&G_{j k}^{\pm}=\delta_{j k}\left(\ell+\sum_{l=1}^{N / 2} \varphi^{+}\left(\lambda_{j}, \lambda_{l}\right)\right)-\varphi^{\pm}\left(\lambda_{j}, \lambda_{k}\right),\\
&\varphi^{\pm}(\lambda, \mu)=\varphi(\lambda-\mu) \pm \varphi(\lambda+\mu),\\
&\bar{f}(\lambda, \mu)=f(\lambda-\mu) f(\lambda+\mu) f(-\lambda-\mu) f(-\lambda+\mu).
\end{align}
\par
Let us first consider the overlap of the crosscap state and an $N$-particle state basis state given by 
\begin{align}
\label{nparticle_LL}
|\boldsymbol{x}_{N}\rangle=&\Phi^{\dagger}\left(x_{1}\right) \ldots \Phi^{\dagger}\left(x_{N}\right)|\Omega\rangle,
\end{align}
From the definition of crosscap state~\eqref{crosscap-0}, we find
\begin{align}
\langle\mathcal{C}|\boldsymbol{x}_{N}\rangle=\prod_{j=1}^{N/2}\delta(x_{j+N/2}-x_{j}-\ell/2).
\end{align}
The overlap of the crosscap state and a Bethe states can be computed as
\begin{align}
\langle\mathcal{C}|\boldsymbol{\lambda}_{N}\rangle=&\int_{T}\mathrm{d}^{N}x\,\left\langle\mathcal{C}|\boldsymbol{x}_{N}\right\rangle\left\langle\boldsymbol{x}_{N}|\boldsymbol{\lambda}_{N}\right\rangle\nonumber\\
=&\int_{T}\mathrm{d}^{N}x\prod_{j=1}^{N/2}\delta(x_{j+N/2}-x_{j}-\ell/2)\chi_{N}\left(\boldsymbol{x}_{N}| \boldsymbol{\lambda}_{N}\right)\nonumber\\
=&\sum_{\sigma\in S_{N}}\prod_{j>k}f(\lambda_{\sigma_j}-\lambda_{\sigma_k})B_{N}(\sigma\boldsymbol{l}_{N}|\ell),
\end{align} 
where the integral region $T$ is defined in~\eqref{domain}, the function $B_{N}(\boldsymbol{l}_{N}|\ell)$ is given by
\begin{align}
B_{N}(\boldsymbol{l}_{N}|\ell)=\left(\prod_{n=1}^{N/2}l_{n}^{\ell/2}\right)\int_{0}^{\ell/2}dx_{N/2}\int_{0}^{x_{N/2}}dx_{N/2-1}\cdots\int_{0}^{x_2}dx_{1}\prod_{j=1}^{N/2}(l_{j}l_{j+N/2})^{x_{j}}.
\end{align}
Following the strategy used in spin chain, we can calculate the overlap. The mainly difference is that the particle position is continuous and the summations become integrals.
\par
For the 2-particle states, one can compute the integral exactly
\begin{align}
B_{2}(\boldsymbol{l}_{2}|\ell)=i a_2^{1/2}\frac{ (a_1a_2)^{1/2}-1}{\lambda_1+\lambda _2}.
\end{align}
The overlap reads
\begin{align}
\langle\mathcal{C}|\boldsymbol{\lambda}_{2}\rangle=i f(l_2,l_1)a_2^{1/2}\frac{ (a_1a_2)^{1/2}-1}{\lambda_1+\lambda _2}+i f(l_1,l_2)a_1^{1/2}\frac{ (a_1a_2)^{1/2}-1}{\lambda_1+\lambda _2},
\end{align}
Similar to the spin chain case, we first take the paired rapidity limit
\begin{align}
\lambda_2\to-\lambda_{1}\,,
\end{align} 
which leads to the result
\begin{align}
\langle\mathcal{C}|\boldsymbol{\lambda}_{2}\rangle=-\ell\sqrt{f(-2\lambda_1)f(2\lambda_1)}=\sqrt{f(-2\lambda_1)f(2\lambda_1)}\det G_{1}^{+}.
\end{align}
For the $N$-particle states, we have 
\begin{align}
B_{N}(\boldsymbol{l}_{N}|\ell)=\sum_{j=0}^{N/2}B_{N,j}(\boldsymbol{l}_{N}|\ell),
\end{align}
where
\begin{align}
B_{N,j}(\boldsymbol{l}_{N}|\ell)=(-1)^{j} \frac{\prod_{k=j+1}^{j+N/2}a_{k}^{1/2}\prod_{k=j+N/2+1}^{N} a_{k}}{\left(\prod_{k=j+1}^{N/2} \sum_{i=j+1}^{k}i\left(\lambda_{i}+\lambda_{i+N/2}\right)\right)\left(\prod_{k=1}^{j} \sum_{i=k}^{j}i\left(\lambda_{i}+\lambda_{i+N/2}\right)\right)}.
\end{align}
By investigating the behavior of $B_{N}(\boldsymbol{l}_{N}|\ell)$ near the pole $\lambda_{m}+\lambda_{m+N/2}=0$, we find that
\begin{align}
B_{N}(\boldsymbol{l}_{N}|\ell)\sim&\frac{\left(a_{m}a_{m+N/2}\right)^{1/2}-1}{i(\lambda_{m}+\lambda_{m+N/2})}\left(a_{m+N/2}\right)^{1/2}B_{N-2,m-1}(\{1,...\cancel{m}...\cancel{m+N/2}...,N\}|\ell).
\end{align}
Interestingly, this relation is exactly the same as the one for Heisenberg spin chain \eqref{eq:mainrecursion}. Following the same steps in spin chain model, one can obtain the overlap 
\begin{align}
\langle\mathcal{C}|\boldsymbol{\lambda}_{N}\rangle&=\prod_{j=1}^{N / 2} \sqrt{f\left(2 \lambda_{j}\right) f\left(-2 \lambda_{j}\right)}\prod_{1 \leq j<k \leq N / 2}\bar{f}\left(\lambda_{j}, \lambda_{k}\right) \times \operatorname{det} G_{N/2}^{+}
\end{align}
Dividing by the norm of Bethe state~\eqref{Bethe-norm-LL}, we finally arrive at the normalized overlap
\begin{align}
\label{normalize-overlap-1}
\frac{\langle\mathcal{C}|\boldsymbol{\lambda}_{N}\rangle}{\sqrt{\langle\boldsymbol{\lambda}_{N}|\boldsymbol{\lambda}_{N}\rangle}}=\sqrt{\frac{\det G_{N/2}^{+}}{\det G_{N/2}^{-}}}.
\end{align} 
We find that indeed the prefactor is again trivial.

\subsection{Crosscap partition function}
\label{sec:3.3}
In~\cite{Caetano:2021dbh}, the crosscap overlaps in IQFTs was obtained by studying the partition function on a cylinder and contract the two ends with the crosscap states, which is related to the Klein bottle partition function. For the Lieb-Liniger model, an analogous quantity is the so-called return amplitude, or Loschmidt amplitude. For a generic initial state, the Loschmidt amplitude is defined by
\begin{align}
\mathscr{L}(\omega)=\langle\Psi_0|e^{-\omega H}|\Psi_0\rangle,
\end{align}
where $\omega$ is a complex number. This quantity plays an important role in the study of quench dynamics and for integrable spin chains it can be computed analytically~\cite{Piroli_2017,Jiang:2021krx}. Now taking $|\Psi_0\rangle=|\mathcal{C}\rangle$ and $\omega=R$ to be a real number, we can define
\begin{align}
\mathscr{L}_{\mathcal{C}}(R)=\langle\mathcal{C}|e^{-RH}|\mathcal{C}\rangle\,.
\end{align}
Using the expansion \eqref{eq:expandC1} and noticing that particle number is conserved, we can equivalently write $\mathscr{L}_{\mathcal{C}}(\omega)$ as
\begin{align}
\label{eq:Loschmidt}
\mathscr{L}_{\mathcal{C}}(R)=\sum_{N=0}^{\infty}\langle\mathcal{C}_{2N}|e^{-RH}|\mathcal{C}_{2N}\rangle=\text{Tr}\,\left[\Pi_c\,e^{-RH}\right],\qquad \Pi_{c}=\sum_{N=0}^{\infty}|\mathcal{C}_{2N}\rangle\langle\mathcal{C}_{2N}|.
\end{align}
The right hand side of \eqref{eq:Loschmidt} takes a very similar form to the Klein bottle partition function of IQFT. Inserting a complete set of Bethe states in \eqref{eq:Loschmidt} and using the exact overlap formula \eqref{normalize-overlap-1}, we obtain
\begin{align}
\mathscr{L}_{\mathcal{C}}(R)=\sum_{N=0}^{\infty}\sum_{\{\boldsymbol{\lambda}_{2N}\}}e^{-R E(\boldsymbol{\lambda}_{2N})}\frac{\det G_{N}^{+}}{\det G_{N}^{-}}.
\end{align}
This is almost the same as a thermal partition function, except for the overlaps. In the thermodynamic limit $\ell\to\infty$, the the ratio of Gaudin-like determinant tends to 1. In this case, the only difference between $\mathscr{L}_{\mathcal{C}}(R)$ and the thermal partition function is that the Bethe roots should be parity even, which agrees exactly with the proposal for crosscap states in IQFT~\cite{Caetano:2021dbh}.\par

In the thermodynamic limit $\ell\to \infty$ and $N\to \infty$ with fixed the particle density $D=N/\ell$, we can compute $\mathscr{L}_{\mathcal{C}}(R)$ by thermodynamic Bethe ansatz (TBA)~\cite{Yang:1968rm}, or the quench action approach~\cite{Caux_2016} in this context. The only modification from the standard TBA is that one should impose the parity even conidition on the Bethe roots. In this limit,  the sum of Bethe roots becomes a path integral over the distribution density
\begin{align}
Z_{2N}\equiv\langle\mathcal{C}_{2N}|e^{-RH}|\mathcal{C}_{2N}\rangle=\int\mathcal{D}[\rho]e^{-2S_o[\rho]+S_{\text{YY}}[\rho]-R E[\rho]+h\ell\left[D-\int_{-\infty}^{\infty}d\lambda\rho(\lambda)\right]}.
\end{align}
The $S_o[\rho]$ comes from the extensive part of the logarithm of the overlaps, or the prefactors in the exact overlap formula. For the crosscap states, however, the prefactor is trivial and hence $S_o[\rho]=0$.
$S_{\text{YY}}[\rho]$ is the Yang-Yang entropy given by
\begin{align}
S_{\text{YY}}[\rho]=\ell\int_{-\infty}^{\infty}d\lambda\left[(\rho+\rho^h)\ln(\rho+\rho^h)-\rho\ln\rho-\rho^h\ln\rho^h\right].
\end{align}
The hole density $\rho^{h}$ is related to the Bethe root density $\rho$ through 
\begin{align}
\label{rho-rho-relation}
\rho(\lambda)+\rho^h(\lambda)=\dfrac{1}{2\pi}+\int_{-\infty}^{\infty}\dfrac{d\mu}{2\pi}\varphi(\lambda-\mu)\rho(\mu),
\end{align}
where the integral kernel is given by~\eqref{LL-phi}. The energy for a given distribution $\rho$ is
\begin{align}
E[\rho]=\ell\int_{-\infty}^{\infty}d\lambda\rho(\lambda)\lambda^2.
\end{align}
The chemical potential $h$ is introduced for the normalization of $\rho(\lambda)$. In addition, one should note the distribution $\rho(\lambda)$ should be an even function because of the parity even condition.
\par
The functional integral can be evaluated by using the saddle point approximation. The saddle point equation is just the Yang-Yang equation 
\begin{align}
\epsilon(\lambda)=\epsilon_0(\lambda)-\frac{1}{R}\int_{-\infty}^{\infty} \frac{d\mu}{2\pi}\varphi(\lambda-\mu)\log\left(1+e^{-R\epsilon(\mu)}\right), 
\end{align}
where
\begin{align}
\frac{\rho(\lambda)}{\rho_{h}(\lambda)}&=e^{-R\epsilon(\lambda)},\quad \epsilon_0(\lambda)=\lambda^2-h.
\end{align}
\par
It is known that distributions are the even functions for the cases of $c\to \infty$ and $c\to 0$~\cite{Yang:1968rm}, which correspond to the free Fermi gas and free Bose gas respectively. These solutions still hold for the crosscap partition functions.
For the limit $R\to 0$, which corresponds to the high temperature limit in the standard TBA, the solution of saddle point equation is given by the uniform distribution 
\begin{align}
\rho(\lambda)=\frac{1}{8\pi},\qquad \rho_{h}(\lambda)=\frac{1}{4\pi}\,.
\end{align}

\section{Dynamical correlation functions in crosscap state}
\label{sec:4}
In the study of out-of-equilibrium dynamics, it is of central importance to compute the time evolution of operators in a given initial state $\langle\Psi_0|\mathcal{O}(t)|\Psi_0\rangle$. Even for integrable models, an exact computation of this quantity is a highly challenging task. For the Lieb-Liniger model, the full time dependence of dynamical correlation functions is only known for very limited cases in the Tonks-Girardeau limit $c\to\infty$~\cite{Collura_2013,PhysRevA.89.013609}.
In this section, we study the dynamical correlation functions in crosscap state of Lieb-Liniger model, namely we compute correlation functions of the form $\langle\mathcal{C}|\mathcal{O}(t)|\mathcal{C}\rangle$. We obtain analytic results in the Tonks-Girardeau limit.\par
In the $c\to\infty$ limit, the model describes the hard-core boson, which behaves like a fermion since the infinite repulsion acts as an effective Pauli principle. Let us denote the hard-core bosonic field as $\tilde{\Phi}(x)$. They obey the usual equal time bosonic commutation relations. The hard-core constraint is imposed by the additional algebraic relations
\begin{align}
[\tilde{\Phi}(x)]^2=[\tilde{\Phi}^{\dagger}(x)]^2=0,\quad \{\tilde{\Phi}(x),\tilde{\Phi}^{\dagger}(x)\}=1.
\end{align}
The non-linear relation between the canonical bosons and hard-core bosons are given by
\begin{align}
\label{eq:projection}
\tilde{\Phi}^{\dagger}(x)=&\mathrm{P}_{x} \Phi^{\dagger}(x) \mathrm{P}_{x}, \quad \mathrm{P}_{x}=|0\rangle\left\langle\left. 0\right|_{x}+| 1\right\rangle\left\langle\left. 1\right|_{x}\right.,
\end{align}
where $\mathrm{P}_{x}$ is the local projector on the truncated Hilbert with at most one boson at $x$. The hard-core bosons are also related to the free fermions via the Jordan-Wigner transformation
\begin{align}
\label{J-W-0}
\Psi(x)=&\exp \left[i \pi \int_{0}^{x} \mathrm{d}z\,\tilde{\Phi}^{\dagger}(z) \tilde{\Phi}(z)\right] \tilde{\Phi}(x),
\end{align}
with the anti-commutation relation
\begin{align}
\{\Psi(x),\Psi(y)\}=\{\Psi^{\dagger}(x),\Psi^{\dagger}(y)\}=0,\quad \{\Psi(x),\Psi^{\dagger}(y)\}=\delta(x-y).
\end{align}
The Jordan-Wigner transformation also maps the hard-core bosonic Hamiltonian to the one describes free fermions. In momentum space, we have  
\begin{align}
H=\sum_{k=-\infty}^{\infty} k^{2}\eta_{k}^{\dagger}\eta_{k}, \quad \eta_{k}=\frac{1}{\sqrt{\ell}}\int_{0}^{\ell}dx e^{-ik x}\Psi(x),\quad k=\frac{2 \pi m}{\ell}.
\end{align}
The time evolution of the fermionic operator is given by~\cite{Collura_2013}
\begin{align}
\eta_{k}(t)=e^{i H t} \eta_{k} e^{-i H t}=e^{-i k^{2} t / 2} \eta_{k}.
\end{align}
Therefore, the time-dependent becomes an overall factor in the Tonks-Girardeau limit, and the dynamical correlation functions can be computed analytically. 
\par
In what follows, we first consider the fermionic two-point and four-point functions. By taking the equal time limit, we obtain the density correlation functions, where the fermionic density is defined as 
\begin{align}
\hat{\rho}(x,t)=\Psi^{\dagger}(x,t)\Psi(x,t).
\end{align}
Note that the bosonic two-point correlation function is different from the fermionic two-point correlation function because it contains an infinite strings of fermionic operators, and the fermionic multi-point function can not factorize into two-point functions functions. However, the fermionic density-density correlation function becomes factorizable in the crosscap state.
\subsection{Two-point function}
\label{sec:4.1}
The two-point function can be expressed as
\begin{align}
\label{eq:2pt}
\langle\Psi^{\dagger}(x_1,t_1)\Psi(x_2,t_2)\rangle_{\mathcal{C}}=&\frac{1}{\ell}\sum_{k_1,k_2}e^{-i(k_1x_1-k_1^2t_1-k_2x_2+k_2^2t_2)}\langle\eta^{\dagger}_{k_1}\eta_{k_2}\rangle_{\mathcal{C}}.
\end{align}
where we used the following notation
\begin{align}
\langle\mathcal{O}\rangle_{\mathcal{C}}=\frac{\langle\mathcal{C}|\mathcal{O}|\mathcal{C}\rangle}{\langle\mathcal{C}|\mathcal{C}\rangle}.
\end{align}
So we need to evaluate the time-independent fermionic two-point correlation
\begin{align}
\label{2-point-eta-c}
\langle\eta^{\dagger}_{k_1}\eta_{k_2}\rangle_{\mathcal{C}}=&\frac{1}{\ell}\int_{0}^{\ell}dz_1dz_2e^{ik_1z_1-ik_2z_2}\langle\Psi^{\dagger}(z_1)\Psi(z_2)\rangle_{\mathcal{C}}.
\end{align}
The fermionic two-point function $\langle\Psi^{\dagger}(z_1)\Psi(z_2)\rangle_{\mathcal{C}}$ can be transformed to the hard-core bosonic correlation function by the Jordan-Wigner transformation~\eqref{J-W-0}. But the crosscap state is defined by the canonical boson operator, which is related to the hard-core boson through the projection \eqref{eq:projection}. Handling the projector in the continuous Lieb-liniger model is somewhat subtle. On the other hand, in the lattice version, the crosscap state is already defined in the truncated Hilbert space. Therefore, we will first calculate the two-point fermionic function in lattice model then take the continuum limit, this strategy was also taken for the computation of correlation functions in the BEC state~\cite{Collura_2013,PhysRevA.89.013609}. 
\par
We consider a one-dimensional lattice model of $L$ sites with lattice spacing $\delta$ and define the lattice operators as follows
\begin{align}
\label{lattice-operators}
b_{m}=\sqrt{\delta} \Phi(m \delta), \quad a_{m}=\sqrt{\delta} \tilde{\Phi}(m \delta), \quad c_{m}=\sqrt{\delta}\Psi(m \delta).
\end{align}
The canonical bosons are denoted as $b_{i},b_{i}^{\dagger}$, which are related to hard-core boson through
\begin{align}
a_{i}=P_ib_{i}P_{i},\quad a_{i}^{\dagger}=P_ib_{i}^{\dagger}P_{i},
\end{align}
where $P_{i}=|0\rangle\langle 0|_{i}+|1\rangle\langle 1|_{i}$ is the on-site projector on the truncated Hilbert space. They satisfy the following (anti)commutation relations
\begin{align}
\left[a_{i}, a_{j}\right] & = [a_{i}^{\dagger}, a_{j}^{\dagger}] = [a_{i}, a_{j}^{\dagger}]  = 0, \quad i \neq j \\
a_{i}^{2} & = (a_{i}^{\dagger})^{2} = 0, \qquad\{a_{i}, a_{i}^{\dagger}\} = 1 .
\end{align}
The Jordan–Wigner transformation in the lattice model becomes 
\begin{align}
c_{i} & = e^{i \pi \sum_{j<i} a_{j}^{\dagger} a_{j}} a_{i} = \prod_{j<i}\left(1-2 a_{j}^{\dagger} a_{j}\right) a_{i},
\end{align}
where the $c_{i},c_{i}^{\dagger}$ are the free fermion operators. Since the crosscap state is already defined in the truncated Hilbert space, we can replace the hard-core boson by the canonical boson. The lattice version of the fermionic two-point function can be calculated using the commutation relations of the bosonic operators, see appendix~\ref{app:C}. The result turns out to be simply
\begin{align}
\langle c_{r}^{\dagger} c_{s}\rangle_{ \mathcal{C}}=\frac{1}{2}\delta_{r,s}.
\end{align}
\par
In the continuum limit, this leads to the two-point time-independent fermionic correlation function
\begin{align}
\langle\Psi^{\dagger}(x)\Psi(y)\rangle_{\mathcal{C}}=\frac{1}{2} \delta(x-y).
\end{align}
Finally, we compute the time dependent two-point correlation function using \eqref{eq:2pt}, where we have taken the limit $\ell\to \infty$, the momentum sum becomes integrals and the final result is
\begin{align}
\langle\Psi^{\dagger}(x_1,t_1)\Psi(x_2,t_2)\rangle_{\mathcal{C}}=&\frac{1}{2\pi}G(x_{12},t_{12}).
\end{align}
Here we have introduced the notation $x_{ij}=x_{i}-x_{j},t_{ij}=t_{i}-t_{j}$. The function $G(x,t)$ is the solution of 1D diffusion equation given by
\begin{align}
G(x,t)=\int_{-\infty}^{\infty} \frac{d k}{2 \pi} e^{-i k x+i k^{2} t}=\frac{1}{2 \sqrt{-i\pi t}} e^{-\frac{ix^{2}}{4 t}}.
\end{align} 
The two-point correlation function is translation invariant, which just depends on the distance between two points. We can also compute the dynamical structure factor as the double Fourier transform of the two-point correlation function in $x_{12}$ and $t_{12}$. The result is simply
\begin{align}
S(p,\omega)=2 \delta \left(\omega -p^2\right).
\end{align}
Apart from the delta function which simply imposes the massive non-relativistic dispersion relation, the result is a constant.
\par
Then the density one-point function can be obtained by taking the limit $t_2\to t_1,x_2\to x_1$, which leads to
\begin{align}
\langle\hat{\rho}(x,t)\rangle_{\mathcal{C}}=\frac{1}{2\pi}\delta(0).
\end{align}
The fermionic density expectation value in crosscap state is time-independent but divergent. This result can be interpreted as follows. The time-independence indicates that the density does not evolve. In fact, applying the quench action approach, we find that the driving term which usually comes from prefactor is vanishing for the crosscap state. Hence the quench action coincides with the Yang-Yang entropy, and the state tends to maximize the entropy, which is given by the uniform distribution. The situation is similar to an infinite temperature thermal partition function. At the same time, the crosscap state is given by a superposition state with arbitrary even number of particles, so that the particle number density is divergent.
\subsection{Four-point function}
\label{sec:4.2}
We then consider the four-point function, which can be written as 
\begin{align}
\label{eq:4pt}
&\langle\Psi^{\dagger}(x_1,t_1)\Psi(x_2,t_2)\Psi^{\dagger}(x_3,t_3)\Psi(x_4,t_4)\rangle_{\mathcal{C}}\nonumber\\
=&\frac{1}{\ell^2} \sum_{k_{1}, k_{2}, k_{3}, k_{4}}e^{-i\left(k_{1}x_1-k_{2}x_2+k_3x_3-k_4x_4-k_{1}^{2}t_1+k_{2}^{2}t_2-k_3^2t_3+k_4^2t_4\right)}\langle\eta_{k_{1}}^{\dagger} \eta_{k_{2}} \eta_{k_{3}}^{\dagger} \eta_{k_{4}}\rangle_{\mathcal{C}}.
\end{align}
So we need to evaluate the initial fermionic four-point correlation function
\begin{align}
\langle\eta_{k_{1}}^{\dagger} \eta_{k_{2}} \eta_{k_{3}}^{\dagger} \eta_{k_{4}}\rangle_{\mathcal{C}}=&\frac{1}{\ell^{2}} \int_{0}^{\ell}\mathrm{d}z_{1}\mathrm{d}z_{2}\mathrm{d}z_{3} \mathrm{d}z_{4}e^{i\left(k_{1} z_{1}-k_{2} z_{2}+k_{3} z_{3}-k_{4} z_{4}\right)}\nonumber\\
&\times \langle\Psi^{\dagger}(z_{1})\Psi(z_{2}) \Psi^{\dagger}(z_{3}) \Psi(z_{4})\rangle_{\mathcal{C}}.
\end{align}
The main ingredient is the four-point function in crosscap state. We first compute the corresponding four-point function in the lattice model using the same technique for the two-point function, then taking the continuum limit. In the lattice model, we need to compute the four-point function $\langle c_{k}^{\dagger} c_{l} c_{m}^{\dagger} c_{j}\rangle_{\mathcal{C}}$. We divide the four operators into two neighbouring pairs and then use Jordan-Wigner transformation on each pairs. The final result can be obtained using the commutation relation of hard-core bosonic operators. We just consider the operators are in the order $k\leq l\leq m\leq j$. For the other orders of $k,l,m,j$, one can also obtain the same formula up to a sign.  The detail calculation is given in appendix~\ref{app:C}, which leads to the result
\begin{align}
\langle c_{k}^{\dagger} c_{l} c_{m}^{\dagger} c_{j}\rangle_{\mathcal{C}}=\frac{1}{4}\delta_{k,l}\delta_{m,j}+\frac{1}{4}\delta_{k,l}\delta_{m,j}\delta_{l,m}+\frac{1}{4}\delta_{k,l}\delta_{m,j}\delta_{l,m-L/2}.
\end{align} 
In the continuum limit, we have
\begin{align}
&\langle\Psi^{\dagger}(z_{1})\Psi(z_{2}) \Psi^{\dagger}(z_{3}) \Psi(z_{4})\rangle_{\mathcal{C}}\nonumber\\
=&\frac{1}{4}\delta(z_2-z_1)\delta(z_4-z_3)\Big(1+\delta(z_3-z_2)+\delta(z_3-z_2-\ell/2)\Big).
\end{align}
Plugging into \eqref{eq:4pt}, we obtain
\begin{align}
\label{Four-point-function}
&\langle\Psi^{\dagger}(x_1,t_1)\Psi(x_2,t_2)\Psi^{\dagger}(x_3,t_3)\Psi(x_4,t_4)\rangle_{\mathcal{C}}\nonumber\\
=&\left(\frac{1}{2\pi}\right)^2G\left(x_{12}, t_{12}\right)G\left(x_{34}, t_{34}\right)\nonumber\\
&+\frac{1}{16\pi}\sqrt{\frac{\tau}{i\pi t_1t_2t_3t_4}}\frac{G\left(x_{23},\frac{\tau}{t_1t_4}\right)G\left(x_{14},\frac{\tau}{t_2t_3}\right)G\left(x_{34},\frac{\tau}{t_1t_2}\right)G\left(x_{12},\frac{\tau}{t_3t_4}\right)}{G\left(x_{24},\frac{\tau}{t_1t_3}\right)G\left(x_{13},\frac{\tau}{t_2t_4}\right)}\nonumber\\
&+\frac{1}{16\pi}\sqrt{\frac{\tau}{i\pi t_1t_2t_3t_4}}\frac{G\left(x_{23}+\frac{\ell}{2},\frac{\tau}{t_1t_4}\right)G\left(x_{14}+\frac{\ell}{2},\frac{\tau}{t_2t_3}\right)G\left(x_{34},\frac{\tau}{t_1t_2}\right)G\left(x_{12},\frac{\tau}{t_3t_4}\right)}{G\left(x_{24}+\frac{\ell}{2},\frac{\tau}{t_1t_3}\right)G\left(x_{13}+\frac{\ell}{2},\frac{\tau}{t_2t_4}\right)},
\end{align}
where we have defined
\begin{align}
\tau=t_1t_2t_{34}+t_{12}t_3t_4.
\end{align}
\par
We are interested in the density-density correlation function,
which can be obtain from the four-point function by taking $t_2\to t_1,t_4\to t_3,x_2\to x_1,x_4\to x_3$. In this case, the last two terms in~\eqref{Four-point-function} become vanishing. The leading order is the first term, the result shows the dynamical density-density correlation function is 
\begin{align}
\langle\rho(x_1,t_1)\rho(x_3,t_3)\rangle_{\mathcal{C}}=\left(\frac{1}{2\pi}\right)^2\delta^2(0),
\end{align}
which is also time-independent but divergent. This result coincides with the product of two fermionic density expectation value in crosscap state, which means the density-density correlation function is factorizable. We learn that the density operator expectation and density-density correlator are time-independent, which is quite different from the general case. For the BEC state, the density-density correlator is time-dependent, but it becomes a stationary one for the late time limit~\cite{PhysRevA.89.013609}.

\section{Conclusion and discussion}
\label{sec:5}
In this paper, we derived a number of new results for crosscap states of both integrable spin chains and the Lieb-Liniger model. For spin chains, we generalized the proof of the integrability of crosscap state to anisotropic and inhomogeneous Heisenberg spin chain. We proved the exact overlap formula of crosscap states and on-shell Bethe states using coordinate Bethe ansatz for both compact and non-compact spin chains.\par

We constructed the crosscap state \eqref{eq:CCstart} for Lieb-Liniger model. This is achieved by taking the proper continuum limits of spin chain models. We provided two methods to derive the crosscap state in Lieb-Liniger model, which lead to the same result. The integrablility of the crosscap state is proved after discretizing the Lieb-Liniger model. The crosscap state has non-vanishing overlaps with any $2N$-particle Bethe states if the Bethe roots are paired. We derived the exact overlap formula of the crosscap state and Bethe states, which is simply given by the ratio of two Gaudin-like determinants, with a trivial prefactor --- the same as in IQFTs and spin chains. These results indicate certain universality of the crosscap overlaps. We considered the Loschmidt echo of the crosscap state. In the thermodynamic limit, it coincides with the thermal partition function with the parity even condition on the Bethe roots. This is similar to the Klein bottle partition function defined for IQFTs.
\par
Moreover, we studied the dynamical fermionic correlation functions in crosscap state in the Tonks-Girardeau limit. We derived analytic results for the two- and four-point functions, which turns out to be surprisingly simple. This gives another valuable data point of analytic results of dynamical correlation functions. By taking the equal time limit, we obtain the density expectation value and the density-density correlator.
We find that the density expectation value is time-independent but divergent and the density-density correlator is factorizable. 
\par
There are some unsolved problems and natural extensions of this work. Firstly, the exact nature of the divergences in the density correlation function should be further clarified. In this regard, it would be interesting to consider the quench dynamics of the crosscap state in the Heisenberg spin chains. One can study the correlation functions for finite chain first and then consider the continuum limit. As usual, the discrete nature of the spin chain would regularize the divergence and provides us useful insights for the continuous model.\par
An intriguing direction for future research would be to extend the concept of the crosscap state to other integrable models which is in the same category, examples include Gaudin-Yang model, Toda chain and Calogero-Sutherland model. One can even consider integrable deformations such as $T\bar{T}$-deformation of these models. The $T\bar{T}$-deformation of the Lieb-Liniger model was studied in~\cite{Cardy:2020olv,Jiang:2020nnb,Hansen:2020hrs}, which turns out to describe dynamical hard rod gas. The study of integrable boundary and boundary states in the $T\bar{T}$-deformed context was initiated in \cite{Jiang:2021jbg}. Our current work gives a clear hint for the construction of crosscap states for these models. It would be interesting to explore the meaning of integrable states in these models and study their properties.\par

Finally, within the Lieb-Liniger model, our construction gives another concrete example of integrable boundary states in addition to the BEC state. It is natural to ask whether one can construct more integrable boundary states for the Lieb-Liniger model. Our current work indicates that considering continuum limit of the spin chain models might be a promising direction.
\section*{Acknowledgments}
We thank Yuan Miao and Ryo Suzuki for helpful discussions.
\appendix
\section{Proof of equation (2.56)}
\label{app:A}
In this appendix, we shall prove that the function $\mathbf{F}(l_m,l_{m+N/2})$ becomes just depend on $\lambda_{m}$ in the paired rapidities limit, namely~\eqref{limit-F}. According to the notation~\eqref{notation-1}, we transform the the arguments $l_{j}$ into $\lambda_{j}$ then consider the limit 
\begin{align}
\lambda_{j+N/2}\to -\lambda_{j},\quad j=1,2,...,N/2.
\end{align}
After this procedure, one can obtain
\begin{align}
\label{eq:A2}
&\prod_{j=m+1}^{m+N/2-1}\left[\frac{f\left(l_{j},l_{m}\right)f\left(l_{m+N/2},l_{j}\right)}{ f\left(l_{m},l_{j}\right)f\left(l_{j},l_{m+N/2}\right)}\right]^{1/2}\nonumber\\
=&\prod_{j=1}^{m-1}\left[\frac{f\left(l_{j+N/2},l_{m}\right)f\left(l_{m+N/2},l_{j+N/2}\right)}{ f\left(l_{m},l_{j+N/2}\right)f\left(l_{j+N/2},l_{m+N/2}\right)}\right]^{1/2}\prod_{j=m+1}^{N/2}\left[\frac{f\left(l_{j},l_{m}\right)f\left(l_{m+N/2},l_{j}\right)}{ f\left(l_{m},l_{j}\right)f\left(l_{j},l_{m+N/2}\right)}\right]^{1/2}\nonumber\\
=&\prod_{j=1}^{m-1}\left[\frac{f\left(-\lambda_{j}-\lambda_{m}\right)f\left(-\lambda_{m}+\lambda_{j}\right)}{ f\left(\lambda_{m}+\lambda_{j}\right)f\left(-\lambda_{j}+\lambda_{m}\right)}\right]^{1/2}\prod_{j=m+1}^{N/2}\left[\frac{f\left(\lambda_{j}-\lambda_{m}\right)f\left(-\lambda_{m}-\lambda_{j}\right)}{ f\left(\lambda_{m}-\lambda_{j}\right)f\left(\lambda_{j}+\lambda_{m}\right)}\right]^{1/2}\nonumber\\
=&\prod_{\substack{j=1 \\ j \neq m}}^{N/2}\left[\frac{f\left(-\lambda_{j}-\lambda_{m}\right)f\left(-\lambda_{m}+\lambda_{j}\right)}{ f\left(\lambda_{m}+\lambda_{j}\right)f\left(-\lambda_{j}+\lambda_{m}\right)}\right]^{1/2}.
\end{align}
Similarly, we have 
\begin{align}
\prod_{j=m+1}^{m+N/2-1}\left[\frac{f\left(l_{j},l_{m+N/2}\right)f\left(l_{m},l_{j}\right)}{f\left(l_{m+N/2},l_{j}\right)f\left(l_{j},l_{m}\right)}\right]^{1/2}=\prod_{\substack{j=1 \\ j \neq m}}^{N/2}\left[\frac{f(\lambda_{j}+\lambda_m)f(\lambda_{m}-\lambda_j)}{f(-\lambda_{m}-\lambda_j)f(\lambda_j-\lambda_{m})}\right]^{1/2}.
\end{align}
In addition, the Bethe equations can be written as
\begin{align}
a_m&=\prod_{\substack{k=1 \\ k \neq m}}^{N}\frac{f(l_k,l_m)}{f(l_m,l_k)}\nonumber\\
&=\prod_{\substack{k=1 \\ k \neq m}}^{N/2}\frac{f(l_k,l_m)f(l_{k+N/2},l_m)}{f(l_m,l_k)f(l_m,l_{k+N/2})}\cdot\frac{f\left(l_{m+N/2}, l_{m}\right)}{f\left(l_{m}, l_{m+N/2}\right)}\nonumber\\
&=\prod_{\substack{k=1 \\ k \neq m}}^{N/2}\frac{f(\lambda_{k}-\lambda_m)f(-\lambda_{k}-\lambda_m)}{f(\lambda_{m}-\lambda_k)f(\lambda_m+\lambda_{k})}\cdot\frac{f\left(-2\lambda_m\right)}{f\left(2\lambda_{m}\right)},
\end{align}
as well as 
\begin{align}
\label{eq:A5}
a_{m+N/2}&=\prod_{\substack{k=1 \\ k \neq m}}^{N/2}\frac{f(\lambda_{k}+\lambda_m)f(\lambda_{m}-\lambda_k)}{f(-\lambda_{m}-\lambda_k)f(\lambda_k-\lambda_{m})}\cdot\frac{f\left(2\lambda_m\right)}{f\left(-2\lambda_{m}\right)}.
\end{align}
Substituting these relations \eqref{eq:A2}-\eqref{eq:A5} into~\eqref{F-factor}, we find most of the terms are cancelled and the final result is
\begin{align}
\mathbf{F}(l_m,l_{m+N/2})=2\sqrt{f(-2\lambda_{m})f(2\lambda_{m})},\quad 1\leq m\leq N/2.
\end{align} 
\section{Deriving the crosscap state in Lieb-Liniger model}
\label{app:B}
We treat more details about deriving the crosscap state in Lieb-Liniger model, which are mentioned in section~\ref{sec:3.1}. 
\par
We first consider the Dyson-Maleev transformation method.
Applying the limit procedure to~\eqref{DM-crosscap}, we arrive at
\begin{align}
|\mathcal{C}\rangle \equiv&\prod_{j=1}^{L/2}\left(1+a^{\dagger}_{j}a^{\dagger}_{j+L/2}\right)|\Omega\rangle\nonumber\\
=&\exp\left[\sum_{j=1}^{L/2}\log\left(1+a^{\dagger}_{j}a^{\dagger}_{j+L/2}\right)\right]|\Omega\rangle\nonumber\\
=&\exp\left[\frac{1}{\delta}\int_{0}^{\ell/2}dx\log\left(1+\frac{1}{L}\left(\sum_{k}e^{-ikx}\tilde{a}^{\dagger}_{k}\right)\left(\sum_{k'}e^{-ik'(x+\ell/2)}\tilde{a}^{\dagger}_{k'}\right)\right)\right]|\Omega\rangle\nonumber\\
=&\exp\left(\frac{1}{\delta}\int_{0}^{\ell/2}dx\log\left(1+\frac{\ell^2}{L(2\pi)^2}\int dk\int dk'e^{-i(k+k')x-ik'\ell/2}\tilde{\Phi}^{\dagger}_{k}\tilde{\Phi}^{\dagger}_{k'}\right)\right]|\Omega\rangle\nonumber\\
\approx&\exp\left[\frac{\ell}{(2\pi)^2}\int_{0}^{\ell/2}dx\int dk\int dk'e^{-i(k+k')x-ik'\ell/2}\tilde{\Phi}^{\dagger}_{k}\tilde{\Phi}^{\dagger}_{k'}\right)|\Omega\rangle\nonumber\\
=&\exp\left(\int_0^{\ell/2} dx{\Phi}^{\dagger}(x){\Phi}^{\dagger}(x+\ell/2)\right)|\Omega\rangle.
\end{align}
In the first step, we rewrite the crosscap into the exponential form. In the second and third step we use the Fourier transformation and take the limiting procedure. In the fourth step, we consider the large $L$ limit but fix the $\ell$. We finally obtain the crosscap state after integrating out the momentum space.
\par
We then consider the discrete version of Lieb-Liniger model, which is closely related to the \emph{generalized} XXX spin chain~\cite{Izergin:1982ry}. By using the definition of the local spin operator~\eqref{Definition-S}, we have
\begin{align}
|\mathcal{C}\rangle \equiv&\prod_{n=1}^{L/2}\left(1+\tilde{S}_{n}^{+}\tilde{S}_{n+\frac{L}{2}}^{+}\right)|\Omega\rangle
=\prod_{n=1}^{L/2}\left(1+\psi_{n}^{\dagger}\psi_{n+L/2}^{\dagger}\right)|\Omega\rangle \nonumber\\
=&\exp\left(\sum_{n=1}^{L/2}\log\left(1+\psi_{n}^{\dagger}\psi_{n+L/2}^{\dagger}\right)\right)|\Omega\rangle\nonumber\\
=&\exp\left(\sum_{n=1}^{L/2}\log\Big(1+\Delta\Phi^{\dagger}(x_n)\Phi^{\dagger}(x_n+\ell/2)\Big)\right)|\Omega\rangle\nonumber\\
\approx&\exp\left(\sum_{n=1}^{L/2}\Delta\Phi^{\dagger}(x_n)\Phi^{\dagger}(x_n+\ell/2)\right)|\Omega\rangle\nonumber\\
=&\exp\left(\int_{0}^{\ell/2}dx\Phi^{\dagger}(x)\Phi^{\dagger}(x+\ell/2)\right)|\Omega\rangle,
\end{align} 
where we use the fact $\psi_{n}$ annihilate the pseudovacuum. 
\par
We can also treat the discrete version of Lieb-Liniger model as non-compact SL(2,R) spin chain then take the continuum limit. The crosscap state defined in SL(2,R) spin chain can be written as
\begin{align}
|\mathcal{C}\rangle =&\prod_{n=1}^{L/2}\exp\left(\tilde{S}_{n}^{+}\tilde{S}_{n+\frac{L}{2}}^{+}\right)|\Omega\rangle\nonumber\\
=&\exp\left(\sum_{n=1}^{L/2}\tilde{S}_{n}^{+}\tilde{S}_{n+\frac{L}{2}}^{+}\right)|\Omega\rangle\nonumber\\
=&\exp\left(\Delta\sum_{j=1}^{L/2}\Phi^{\dagger}(x_n)\Phi^{\dagger}(x_n+\ell/2)\right)|\Omega\rangle\nonumber\\
=&\exp\left(\int_{0}^{\ell/2}dx\Phi^{\dagger}(x)\Phi^{\dagger}(x+\ell/2)\right)|\Omega\rangle.
\end{align}
\par
Finally, we arrive at the same crosscap state in Lieb-Liniger model by taking the continuum limit from both Heisenberg spin chains and non-compact SL(2,R) spin chain.
\section{Lattice computation of correlation functions}
\label{app:C}
This section is the detail calculation of the correlation function in lattice model. Since the crosscap state is already defined in truncated Hilbert space, we can straightforward obtain the time-independent correlation function. According to the definition of the crosscap state and the commutation relation of the hard-core boson $a_{j}$, we find the following relations
\begin{align}
&a_{r}^{\dagger}a_{s}|\mathcal{C}\rangle=a_{r}^{\dagger}a_{s+L/2}^{\dagger}\prod_{j=1,j\ne s}^{L/2}\left(1+a^{\dagger}_{j}a^{\dagger}_{j+L/2}\right)|\Omega\rangle,\quad 1\leq s<L/2,\\
&a_{r}^{\dagger}a_{s+L/2}|\mathcal{C}\rangle=a_{r}^{\dagger}a_{s}^{\dagger}\prod_{j=1,j\ne s}^{L/2}\left(1+a^{\dagger}_{j}a^{\dagger}_{j+L/2}\right)|\Omega\rangle,\quad 1\leq s<L/2,\\
&a_{r}^{\dagger}a_{s}^{\dagger}|\mathcal{C}\rangle=a_{r}^{\dagger}a_{s}^{\dagger}\prod_{j=1,j\ne s}^{L/2}\left(1+a^{\dagger}_{j}a^{\dagger}_{j+L/2}\right)|\Omega\rangle,\quad 1\leq s<L/2,\\
&a_{r}^{\dagger}a_{s+L/2}^{\dagger}|\mathcal{C}\rangle=a_{r}^{\dagger}a_{s+L/2}^{\dagger}\prod_{j=1,j\ne s}^{L/2}\left(1+a^{\dagger}_{j}a^{\dagger}_{j+L/2}\right)|\Omega\rangle,\quad 1\leq s<L/2,
\end{align}
which lead to (for $r\leq s$)
\begin{align}
&\langle\mathcal{C}|a^{\dagger}_{r}a_{s}|\mathcal{C}\rangle=2^{L/2-1}\delta_{r,s},\\
&\langle\mathcal{C}|a_{r}a^{\dagger}_{s}|\mathcal{C}\rangle=2^{L/2-1}\delta_{r,s},\\
&\langle\mathcal{C}|a^{\dagger}_{r}a^{\dagger}_{s}|\mathcal{C}\rangle=2^{L/2-1}\delta_{s,r+L/2},\\
&\langle\mathcal{C}|a_{r}a_{s}|\mathcal{C}\rangle=2^{L/2-1}\delta_{s,r+L/2},
\end{align}
Notice the norm of the crosscap state is $\langle\mathcal{C}|\mathcal{C}\rangle=2^{L/2}$, we then obtain that (for $r\leq s$)
\begin{align}
\frac{\langle\mathcal{C}|c_{r}^{\dagger} c_{s}| \mathcal{C}\rangle}{\langle\mathcal{C}|\mathcal{C}\rangle}&=\frac{\langle\mathcal{C}|a_{r}^{\dagger} \prod_{p=r+1}^{s-1}\left(1-2 a_{p}^{\dagger} a_{p}\right) a_{s}| \mathcal{C}\rangle}{\langle\mathcal{C}|\mathcal{C}\rangle}=\frac{1}{2} \delta_{r,s},\\
\frac{\langle\mathcal{C}|c_{r} c^{\dagger}_{s}| \mathcal{C}\rangle}{\langle\mathcal{C}|\mathcal{C}\rangle}&=-\frac{\langle\mathcal{C}|a_{r} \prod_{p=r+1}^{s-1}\left(1-2 a_{p}^{\dagger} a_{p}\right) a^{\dagger}_{s}| \mathcal{C}\rangle}{\langle\mathcal{C}|\mathcal{C}\rangle}=-\frac{1}{2} \delta_{r,s},\\
\frac{\langle\mathcal{C}|c^{\dagger}_{r} c^{\dagger}_{s}| \mathcal{C}\rangle}{\langle\mathcal{C}|\mathcal{C}\rangle}
&=\frac{\delta_{r+L/2,s}\langle\mathcal{C}|a^{\dagger}_{r}a^{\dagger}_{r+L/2} \prod_{p=1,p\neq r}^{L/2}\left(1-2 a_{p}^{\dagger} a_{p}\right)|\mathcal{C}\rangle}{\langle\mathcal{C}|\mathcal{C}\rangle}=\frac{1}{2}\delta_{r+L/2,s}\delta_{L,2},\\
\frac{\langle\mathcal{C}|c_{r} c_{s}| \mathcal{C}\rangle}{\langle\mathcal{C}|\mathcal{C}\rangle}
&=-\frac{\delta_{r+L/2,s}\langle\mathcal{C}|a_{r}a_{r+L/2} \prod_{p=1,p\neq r}^{L/2}\left(1-2 a_{p}^{\dagger} a_{p}\right)|\mathcal{C}\rangle}{\langle\mathcal{C}|\mathcal{C}\rangle}=-\frac{1}{2}\delta_{r+L/2,s}\delta_{L,2}.
\end{align}
where we use the Jordan-Wigner transformation. If we consider the length of the chain larger than 2, the last two cases become vanishing. 
\par
For the four-point function, we first consider the indices are in the order $k\leq l\leq m\leq j$
\begin{align}
\langle\mathcal{C}|c_{k}^{\dagger} c_{l} c_{m}^{\dagger} c_{j}|\mathcal{C}\rangle
\end{align} 
We split the four operator into two pairs, then use the Jordan-Wigner transformation for each pair. The non-vanishing of the correlation indices must be paired $k=l,m=j$, namely
\begin{align}
\langle c_{l}^{\dagger} c_{l} c_{j}^{\dagger} c_{j}\rangle_{\mathcal{C}}=\left\{\begin{array}{ll}
1 / 4 & i \neq j, j+\frac{L}{2} \\
1/2 & i=j, j+\frac{L}{2}
\end{array}\right..
\end{align}
\par
If the operators are not in this order $k\leq l\leq m\leq j$, we permute them so that they are spatially ordered. Note that the final order of the two creation and the two annihilation operators may be different from the original one. There are three situations
\begin{align}
c_{k} c_{l}&=a_{k} \prod_{p=k}^{l-1}\left(1-2 a_{p}^{\dagger} a_{p}\right) a_{l}=-a_{k} \prod_{p=k+1}^{l-1}\left(1-2 a_{p}^{\dagger} a_{p}\right) a_{l},\\
c_{k} c_{l}^{\dagger}&=a_{k} \prod_{p=k}^{l-1}\left(1-2 a_{p}^{\dagger} a_{p}\right) a_{l}^{\dagger}=-a_{k} \prod_{p=k+1}^{l-1}\left(1-2 a_{p}^{\dagger} a_{p}\right) a_{l}^{\dagger},\\
c_{k}^{\dagger} c_{l}^{\dagger}&=a_{k}^{\dagger} \prod_{p=k}^{l-1}\left(1-2 a_{p}^{\dagger} a_{p}\right) a_{l}^{\dagger}=a_{k}^{\dagger} \prod_{p=k+1}^{l-1}\left(1-2 a_{p}^{\dagger} a_{p}\right) a_{l}^{\dagger},
\end{align}
which follow the results
\par
\begin{center}
\begin{tabular}{|p{2cm}||p{2cm}|p{2cm}|p{2cm}|p{2cm}|p{2cm}|}
 \hline
 operator & 0 pair &1 pair  & 2 pairs & 2 pairs\par$(l=m)$ & 2 pairs \par$(l+\frac{L}{2}=m)$\\
 \hline
$\langle c_{k}^{\dagger} c^{\dagger}_{l} c_{m} c_{j}\rangle_{\mathcal{C}}$ & 0 & 0 & 0 & 0&0\\&&&&& \\
 $\langle c_{k}^{\dagger} c_{l} c^{\dagger}_{m} c_{j}\rangle_{\mathcal{C}}$ &0&0&$\frac{1}{4}$&$\frac{1}{2}$&$\frac{1}{2}$\\&&&&& \\
 $\langle c_{k}^{\dagger} c_{l} c_{m} c^{\dagger}_{j}\rangle_{\mathcal{C}}$ &0&0&$-\frac{1}{4}$&0&0\\&&&&& \\
 $\langle c_{k} c^{\dagger}_{l} c^{\dagger}_{m} c_{j}\rangle_{\mathcal{C}}$ &0&0&$-\frac{1}{4}$&0&0\\&&&&& \\
 $\langle c_{k} c^{\dagger}_{l} c_{m} c^{\dagger}_{j}\rangle_{\mathcal{C}}$ &0&0&$\frac{1}{4}$&$\frac{1}{2}$&$\frac{1}{2}$\\&&&&& \\
 $\langle c_{k} c_{l} c^{\dagger}_{m} c^{\dagger}_{j}\rangle_{\mathcal{C}}$ &0&0&0&0&0\\
 \hline
\end{tabular}
\end{center}
We conclude the results as
\begin{align}
\langle c_{k}^{\dagger} c_{l} c^{\dagger}_{m} c_{j}\rangle_{\mathcal{C}}=(-1)^{\omega}\frac{1}{4}\delta_{k,l}\delta_{m,j}\Big(1+(-1)^{\omega}(\delta_{l,m}+\delta_{l+M/2,m})\Big),
\end{align}
where the $\omega$ represents the number of pairs beginning with annihilation operator.
\bibliographystyle{JHEP}
\bibliography{reference}
\end{document}